\documentclass[a4paper,11pt]{article}
\pdfoutput=1 

\usepackage{jheppub2} 
\usepackage{bbm,bm,graphicx,mathtools,color,hyperref,slashed}
\usepackage{amssymb,amsmath}
\usepackage{extarrows}
\usepackage{comment}
\usepackage[all]{xy}
\usepackage[dvipsnames]{xcolor}

\graphicspath{{./Figures/}}

\newcommand{\diff}{\mathrm{d}}

\newcommand{\tr}{\mathrm{tr}}

\newcommand{\im}{\mathrm{i}}

\newcommand{\rme}{\mathrm{e}}

\newcommand*{\half}{\textstyle{ \frac{1}{2}}}
\def\Z{{\mathbb Z}}
\def\R{{\mathbb R}}

\preprint{.}

\title{The metamorphosis of semi-classical mechanisms of 
confinement: From monopoles on  ${\mathbb R}^3 \times S^1$ 
to  center-vortices  on  ${\mathbb R}^2 \times T^2$ }

\author[1]{Canberk G\"uvendik,}
\author[1]{Thomas Schaefer,} 
\author[1]{Mithat \"Unsal}

\affiliation[1]{Department of Physics, North Carolina State University, Raleigh, NC 27607, USA}

\abstract{There are two distinct regimes of Yang-Mills theory where we 
can demonstrate confinement, the existence of a mass gap, and the  
multi-branch structure of the effective potential as a function of
the theta angle using a reliable semi-classical calculation.
The two regimes are deformed Yang-Mills theory on ${\mathbb R}^3
\times S^1$, and Yang-Mills theory on ${\mathbb R}^2\times T^2$ 
where the torus is threaded by a 't Hooft flux. The weak coupling 
regime is ensured by the small size of the circle or torus. In the 
first case the confinement mechanism is related to self-dual 
monopoles, whereas in the second case self-dual center-vortices 
play a crucial role. These two topological objects are distinct. 
In particular, they have different mutual statistics with Wilson 
loops. On the other hand, they carry the same topological charge 
and action. We consider the theory on ${\mathbb R \times T^2 
\times S^1}$ and extrapolate both the monopole and vortex regimes 
to a quantum mechanical domain, where a cross-over takes place. Both 
sides of the cross-over are described by a deformed $\Z_N$ TQFT. On
${\mathbb R^2 \times S^1 \times S^1}$, we derive an effective field 
theory (EFT) of vortices from the EFT of monopoles in the presence 
of a 't Hooft flux. This construction is based on a two-stage Higgs
mechanism, reducing $SU(N)$ to $U(1)^{N-1}$ in 3d first, followed
by reduction to a $\Z_N$ EFT in 2d in the second step. This result
shows how monopoles transmute into center-vortices, and suggests 
adiabatic continuity between the two confinement mechanisms. The 
basic mechanism is flux fractionalization: The magnetic flux of the
monopoles splits up and is collimated in such a way that 2d Wilson 
loops detect it as a center vortex.  }

\begin{document}
\maketitle

\section{Overview}\label{sec:introduction}

 Pure Yang-Mills theory as well as QCD(adj) with massless 
or massive fermions in the adjoint representation of the 
gauge group are theories in which there exists a $\Z_N^{[1]}$
1-form symmetry, known as center-symmetry. In these theories
the notion of confinement is sharply defined, and the Wilson
loop is a true order parameter. Both pure Yang-Mills theory
and QCD(adj) admit a semi-classical weak coupling description
for (non-thermal) compactifications of the theory on 
four-manifolds such as $\R^3 \times S^1$\cite{Unsal:2007vu,
Unsal:2007jx,Unsal:2008ch,Shifman:2008ja,Davies:2000nw,
Poppitz:2021cxe} and $\R^2 \times T^2$ \cite{Tanizaki:2022ngt,
Tanizaki:2022plm, Hayashi:2023wwi, Hayashi:2024qkm}. These 
realizations are adiabatically connected to the theory on 
$\R^4$. In the limit where the size of the circle or the 
torus are small non-perturbative effects such as confinement,
dependence on the theta angle, and the mass gap can be 
understood analytically. 

 An important challenge that arises from these studies is 
the precise relationship between semi-classical confinement on 
$\R^3\times S^1$ and $\R^2\times T^2$ \cite{Tanizaki:2022ngt}. 
On $\R^3 \times S^1$ we can select the center-symmetric phase
using either specific boundary conditions for fermions or 
double-trace deformations, and confinement is generated by 
monopole instantons or magnetic bions \cite{Unsal:2007vu,
Unsal:2007jx,Unsal:2008ch,Shifman:2008ja, Poppitz:2012sw, 
Davies:2000nw}. On $\R^2 \times T^2$ the center symmetric 
phase is stabilized by a 't Hooft flux, and it was shown that
confinement is caused by center vortices on $\R^2$ in a manner
reminiscent of the charge-$N$ abelian Higgs model in 2d 
\cite{Tanizaki:2022ngt,Tanizaki:2022plm, Hayashi:2023wwi,
Hayashi:2024qkm, Hayashi:2024gxv}, see also \cite{Poppitz:2022rxv,
Cox:2021vsa,Anber:2023sjn,Yamazaki:2017dra, Yamazaki:2017ulc}.

 There are some obvious commonalities between the two 
scenarios. Both monopole instantons and center-vortices can 
be viewed (in certain circumstances) as configurations 
with fractional topological charge and fractional action
\begin{align}
Q_T= \frac{1}{N},  \qquad S_0 = \frac{1}{N} S_I  
  = \frac{8\pi^2}{g^2N}, 
\end{align}
where $S_I$ is the 4d instanton action \cite{Kraan:1998pm,
Kraan:1998sn,Lee:1998bb,Lee:1997vp}. However, these are
also important differences. In particular, center-vortices
have non-trivial mutual statistics with the Wilson loop 
\cite{Gonzalez-Arroyo:1998hjb, Montero:1999by,Montero:2000pb,
Gonzalez-Arroyo:2023kqv}, while monopoles do not. More 
specifically, monopoles on $\R^3 \times S^1$ appear 
when the Polyakov loop around $S^1$ acquires a non-trivial
vev $P_3 \propto C$, where $C$ is clock matrix. This leads 
to dynamical abelianization, $SU(N) \rightarrow
U(1)^{N-1}$. In contrast, on $\R^2 \times T^2$, center-vortices
appear when the Polyakov loops on $T^2$ cycles are non-commuting
pairs, $P_1=C, P_2 = S$, such as clock and shift matrix, and 
the gauge structure is reduced down to the center, $SU(N)
\rightarrow \Z_N$. The two scenarios can be summarized as 
follows: 
\begin{align} 
&\R^3 \times S^1: \qquad 
  SU(N)\; \xrightarrow{P_3\;\text{adjoint Higgs }}\;  
      U(1)^{N-1}
  \qquad   { \rm monopoles, \;  bions,}  \cr   
&\R^2 \times T^2: \qquad  
  SU(N)\; \xrightarrow{P_{1,2}\;\text{adjoint Higgs}}\;
    \Z_N   \qquad  \qquad  
      { \rm center \; vortices}\, .   
\end{align} 

\begin{figure}[tbp] %
\begin{center}
\includegraphics[width=1\textwidth]{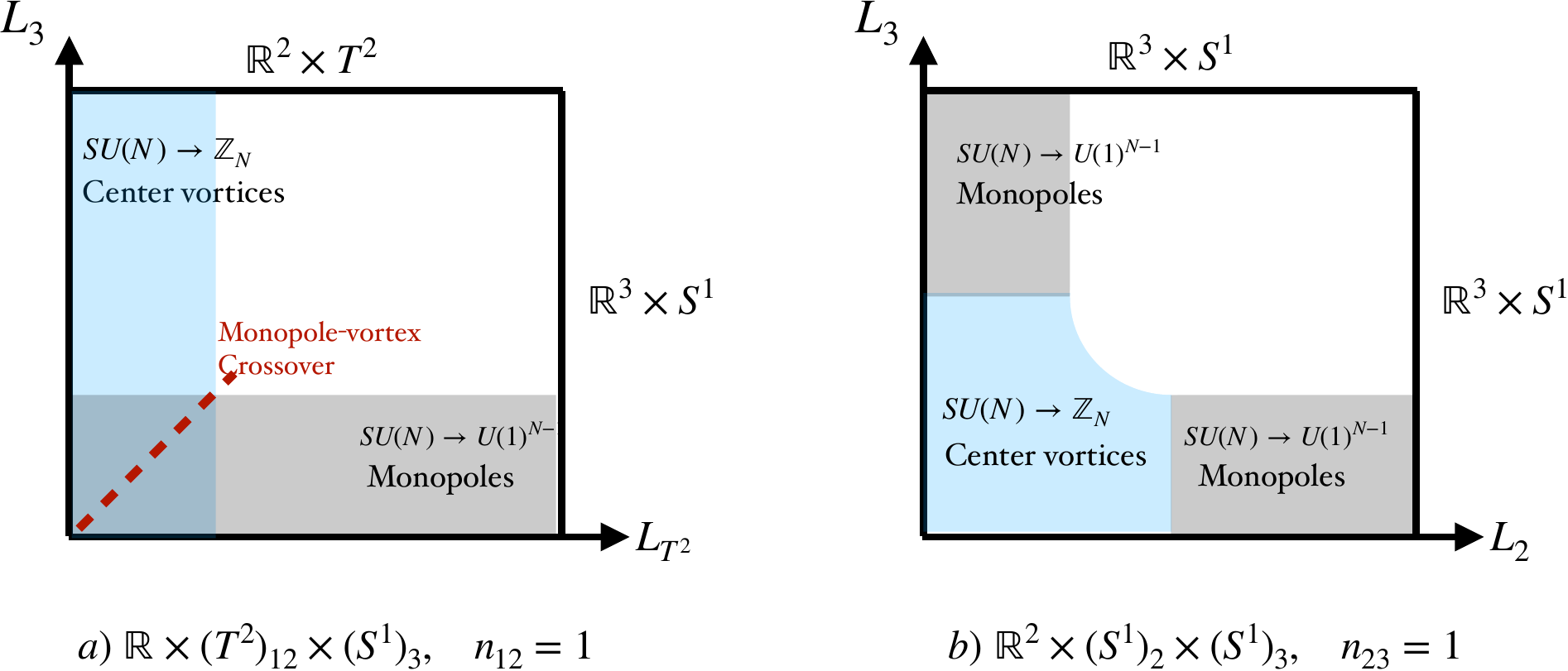} 
\end{center}
\caption{a) Semi-classical regimes of $SU(N)$ gauge theory 
on $\R \times T^2 \times S^1$ with one unit of 't Hooft flux
$n_{12}=1$. For $ {\rm min}(L_{S^1}, L_{T^2}) \ll \Lambda^{-1} $, 
the theory is amenable to semi-classical treatment. The gauge 
structure is reduced to $U(1)^{N-1}$ by a Higgs mechanism in 
the monopole regime (shaded gray), and  to $\Z_N$ in center-vortex 
regime (shaded light blue). The two regimes are adiabatically 
connected, with a cross-over in between along the red line.  
b) Semi-classical regimes $\R^2 \times S^1 \times S^1$ with  
$n_{23}=1$  units of 't Hooft flux \cite{Hayashi:2024yjc}. The theory undergoes a
two-stage Higgs mechanism. In the first stage, monopole mechanism is operative and after the second, center-vortices take over.}
\label{fig:phases3d}
\end{figure}

  The goal of the present work is to make the connection
between these two regimes more precise. For this purpose
we will consider two compactifications that interpolate
between the monopole and the vortex scenario, Yang-Mills 
theory on $\R\times T^2\times S^1$ as well as $\R^2\times
T^2=\R^2\times S^1\times S^1$, see Fig.\ref{fig:phases3d}. 
We will argue that the two semi-classical regimes are 
adiabatically connected through a monopole-vortex crossover, 
see the lower left corner of \ref{fig:phases3d}a) and 
\S.\ref{sec:sym} and \S.\ref{sec:T2small}. We will also 
show how vortices can arise from monopoles in the presence 
of a 't Hooft flux, see Fig.\ref{fig:phases3d}b) and 
\S.\ref{sec:asym}.

 These results are remarkable, because they provide a unified
perspective of monopole-induced confinement on $\R^3 \times S^1$ 
and center-vortex-induced confinement on $\R^2 \times T^2$, and their relation to 4d instantons.
Historically, the monopole and center vortex pictures of
confinement in QCD were viewed as distinct alternatives for 
the confinement mechanism on $\R^4$ \cite{tHooft:1977nqb,
Cornwall:1979hz,Nielsen:1979xu}. In the present work we
demonstrate that both mechanisms are realized in different 
semi-classical regimes of the Yang-Mills theory which are
adiabatically connected to $\R^4$. Furthermore, both monopole and center-vortex 
arise, from  physical fractionalization of 4d instanton, by either non-trivial gauge holonomy or 't Hooft flux.

 In this paper we also study the connection between 
the effective theories (EFTs) of the monople and vortex
scenarios. The long distance theory on $\R^3 \times S^1$ 
is an EFT based on the grand canonical description of a 
magnetic Coulomb gas \cite{Unsal:2008ch}, similar to  the 
Polyakov model on $\R^3$ \cite{ Polyakov:1975rs,Polyakov:1987ez}.
We will refer to this theory as the monopole EFT. In the 
long-distance regime on  $\R  \times T^2 \times S^1$, and 
$\R^2 \times T^2$, the EFT is a $\Z_N$ topological quantum 
field theory (TQFT) \cite{Banks:2010zn,Kapustin:2014gua} 
deformed by local topological operators \cite{Nguyen:2024ikq, 
Cherman:2021nox} in 1d and 2d, respectively. The local
topological operators are center-vortices in 2d and fractional
instantons in 1d. We will study the transition between 
these effective field theories,
\begin{align}
  {\rm Monopole \; EFT \;  in \; 3d} \Longrightarrow   {\rm TQFT 
  + deformation \; in \; 1d \; and \; 2d}   \, . 
\end{align}
The two regimes meet at a quantum mechanical domain, where 
a cross-over and dramatic rearrangement of states occur,  see
\S.\ref{Sec:two-hierarchies}.

 In \S.\ref{sec:asym} we discuss $\R^2 \times T^2$. The most
remarkable finding is the transmutation of monopoles and the 
attached magnetic flux into center-vortices.\footnote{
In $d>2$ dimensions, most of the arguments in favor of the 
center-vortex mechanism are based on numerical studies 
\cite{DelDebbio:1996lih,DelDebbio:1998luz,Engelhardt:1998wu,
Alexandrou:1999iy,Alexandrou:1999vx,deForcrand:1999our,
deForcrand:2000pg,Sale:2022qfn}. In the center-vortex literature,
there are also models which view monopoles as junctions of 
center-vortices, see e.g.\cite{Greensite:2003bk, Deldar:2015kga, 
Oxman:2018dzp}.}  
Consider a monopole labeled by one of the roots $\alpha_i$
of the gauge group. At  short distances the total magnetic 
flux $2 \pi\alpha_i$ is spherically symmetric. We show that 
in the presence of a 't Hooft flux \cite{tHooft:1979rtg,
tHooft:1981sps, vanBaal:1982ag} the magnetic flux 
fractionalizes and collimates into tubes of finite thickness. 
In the 2d EFT a large Wilson loop which encircles the flux 
tube sees it as center-vortex. In particular, the relationship
is topological and the Wilson loop acquires a phase $W(C) = 
e^{- \im \frac{2\pi}{N}}$ if the vortex threads $C$, and 
$W(C) =1$ if it does not. In contrast, in the 3d EFT, the 
Wilson loop acquires a phase controlled by a geometric 
quantity, the solid angle $\Omega_D(x)$ subtended by the 
oriented surface $D$ $(\partial D=C)$ relative to the 
location of the monopole. We provide a detailed explanation 
of this phenomenon, which we refer refer to  as \emph{flux
fractionalization} in \S.\ref{sec:fluxfrac}. We also derive 
a 2d EFT in terms of the degrees of freedom of the 3d EFT, 
the dual photons.


\subsection{The set-up}

 Throughout the paper, we study $4$-dimensional $SU(N)$ Yang-Mills
(YM) theory on a $4$d spacetime  $M_4$. The classical action of
the theory is given by 
\begin{align}
S_{\mathrm{YM}}={1\over g^2}\int \tr[F(a)\wedge \star F(a)]
  +{\im\, \theta\over 8\pi^2}\int \tr[F(a)\wedge F(a)] \, , 
\label{classicalaction}
\end{align}
where $a$ is the $SU(N)$ gauge field, $F(a)=\diff a + \im 
a\wedge a$ is the field strength, $\theta$ is the topological
angle, and ${1\over 8\pi^2}\int \tr[F(a)\wedge F(a)]\in \Z$ 
is the instanton number. The theory has a $\mathbb{Z}_N$ 
$1$-form symmetry \cite{Gaiotto:2014kfa} (known as 
center-symmetry) denoted by $\mathbb{Z}_N^{[1]}$. The 
4-manifolds we consider are 
\begin{align} 
M_4 = \mathbb{R} \times  (S^1)_1 \times (S^1)_2 \times (S^1)_3
 &\quad  {\rm with } 
    \quad  (x_0, x_1, x_2, x_3)\in M_4, 
\label{setup} 
\end{align}
with circle sizes $L_1, L_2, L_3$. This includes the limits 
$\R^3 \times S^1$ and $\R^2 \times T^2$. In  \S.\ref{sec:sym} 
and \S.\ref{sec:T2small}, we turn on $n_{12} =1$ units of 
't Hooft flux, and work with a symmetric $T^2$ with $L_1=L_2=
L_{T^2}$.  We explore different semi-classical regimes by
changing $L_{T^2}/L_3$. In \S.\ref{sec:asym}, we work with
$n_{23} =1$ on $\R^2 \times  (S^1)_2 \times (S^1)_3$ and 
an asymmetric torus, and explore the dynamics by adjusting 
$L_2/L_3$.

\section{Semi-classical analysis on $\R \times T^2_{\rm large}
\times S^1_{\rm small} $ with a 't Hooft flux through the 
torus $T^2$}
\label{sec:sym} 


 We consider compactifications on the 4-manifold 
\begin{align} 
M_4 &= \mathbb{R}\times  \underbrace{(T^2)_{12}}_{n_{12} =1 \;  
{\rm  flux}} \times (S^1)_3 \, .  
\label{setup1} 
\end{align}
The large-$T^2$ and large-$S^1$ limits correspond to the 
limiting cases
\begin{align} 
 \R \times T^2_{\rm large} \times S^1_{\rm small} \sim  
 \R^3 \times S^1,  
\label{setup1-m}  \\
 \R \times T^2_{\rm small} \times S^1_{\rm large}  
  \sim  \R^2 \times T^2\, . 
\label{setup1-v}
\end{align}
In order to study the theory in the weak coupling limit 
we will assume that 
\begin{align} 
 {\rm min}( L_{S^1}, L_{T^2})  \ll \Lambda^{-1}\, ,
\end{align}
where $\Lambda^{-1}$ is the strong coupling scale of the 
Yang-Mills. In this section we will consider scenario
(\ref{setup1-m}), which corresponds to the faint blue 
area along the $y$-axis of Fig.~\ref{fig:phases3d}a).
We will study the gray band along the $x$-axis in 
\S.\ref{sec:T2small}, and describe the crossover regime
in \S.\ref{Sec:two-hierarchies}.

\subsection{'t Hooft flux: Alternative implementations and 
energetics}
 
 A 't Hooft flux on $T^2$ is characterized by the gauge invariant
integers $n_{12} \in \Z_N$ that appear in the twisted boundary
conditions. The identification of fields  at $(x_1,x_2)$, 
$(x_1+L_1, x_2)$  and $(x_1,x_2+L_2)$ is achieved by using the 
transition functions $g_{1}(x_2), g_2(x_1)$.  
\begin{align}
 & a(x_1=L_1,x_2)=g_1(x_2)^{\dagger}a(x_1=0,x_2)g_1(x_2)-
    \im g_1(x_2)^{\dagger}\diff g_1(x_2), \nonumber \\
 & a(x_1,x_2=L_2)=g_2(x_1)^{\dagger}a(x_1,x_2=0)g_2(x_1)-
    \im g_2(x_1)^{\dagger}\diff g_2(x_1)\, .
\end{align}
The consistency at the corners  demands the cocycle condition ~\cite{tHooft:1979rtg} 
\begin{equation}
g_1(L_2)^{\dagger}g_2(0)^{\dagger}=g_2(L_1)^{\dagger}
g_1(0)^{\dagger}\exp
\left( \im { \frac{2\pi}{ N}}n_{12}\right). 
    \label{eq:tHooft_twist_continuum}
\end{equation}
The label $n_{12} \in\mathbb{Z}_N$ is the gauge invariant data in the transition functions and is called  't~Hooft flux.   

The 't~Hooft flux may also be viewed as a background gauge field for the $\Z_N^{[1]}$ 
1-form symmetry \cite{Kapustin:2014gua}. This is usually achieved by 
introducing a pair of $U(1)$ 2-form and 1-form gauge fields $(B^{(2)},B^{(1)})$ satisfying \begin{align}
    N B^{(2)}= \diff B^{(1)}, \qquad \int B^{(2)} \in \frac{2 \pi}{N}\Z,   \qquad {\rm or} \qquad   B^{(2)} = \frac{2 \pi n_{12}}{N L_1 L_2x}  \diff x_1 \wedge \diff x_2 
    \label{two-form}
\end{align}

\begin{figure}[t]
\begin{center}
 \includegraphics[angle=0, width=0.8\textwidth]{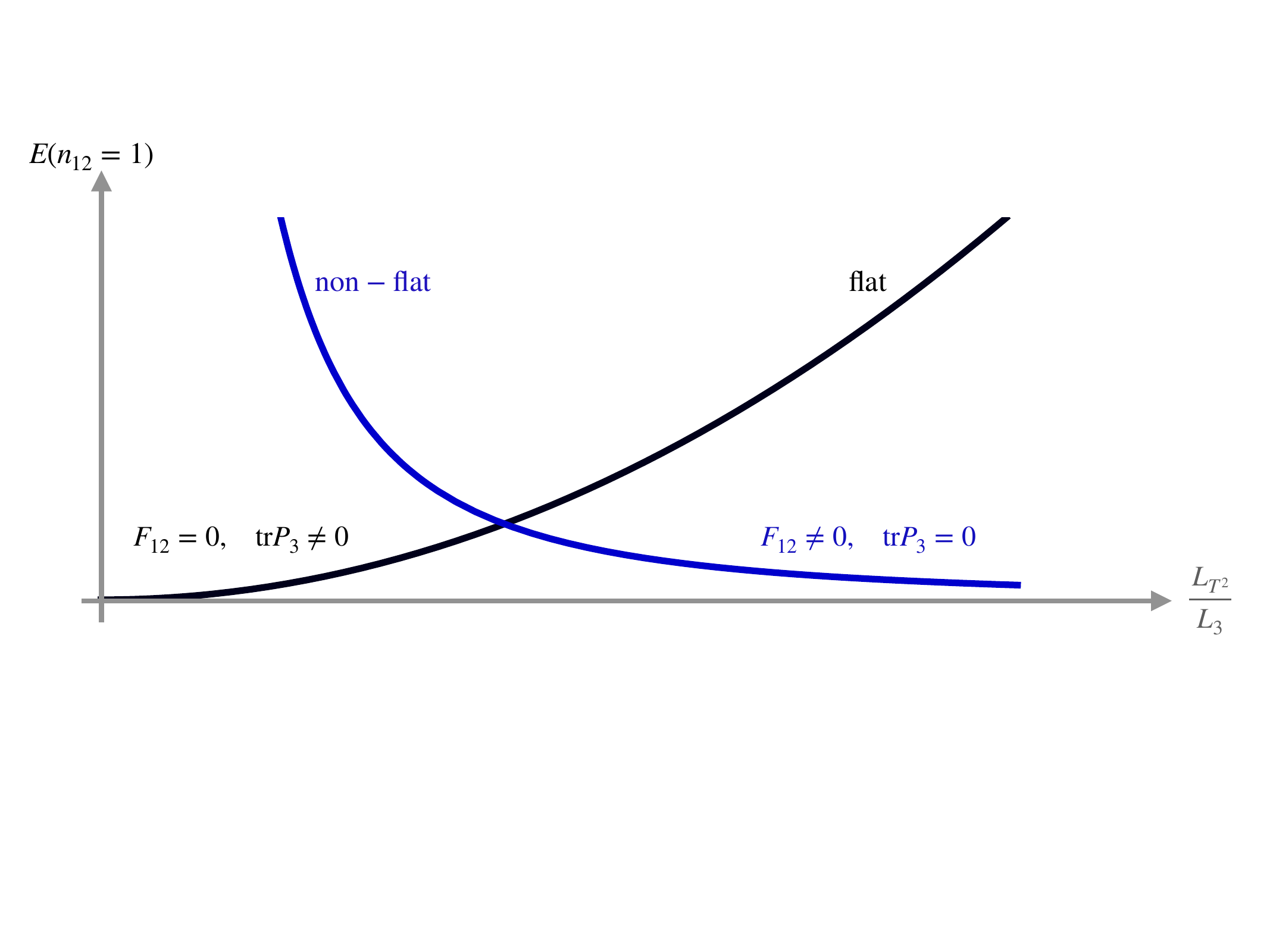}
\vspace{-3cm}
\caption{Energy of flat and non-flat background chromo-magnetic fields in the $n_{12}=1$ 't Hooft flux sector. 
For $L_{T^2} \gg L_3$, the ground state is a configuration with $F_{12} = \frac{2 \pi}{L_1 L_2} H^{(i)} $.   For 
$L_{T^2} \ll L_3$,  the ground state is  a flat connection, $F_{12} =0$.
 }
\label{compete}
\end{center}
\end{figure}

\noindent
{\bf Does $F_{12}$ vanish in the ground state?}  Turning on a 't~Hooft flux $n_{12} \neq 0$   is a kinematic construction.  
There are different gauge choices for the transition functions  \eqref{eq:tHooft_twist_continuum} that produce   
 $n_{12} \neq 0$. 
 Two convenient choices are   non-abelian constant matrices  and abelian space dependent transition matrices.  These are called $\Gamma$-gauge and $\Omega$-gauge, respectively,  in \cite{Poppitz:2022rxv},  and are given 
 by 
 \begin{align} 
& \Gamma- {\rm gauge}: \;\; g_1(x_2)    = S,  \qquad  g_2(x_1)     = C \\
 &\Omega- {\rm gauge}: \;\;  g_1(x_2)    = 1,  \qquad  g_2(x_1)     = e^{i \frac{2 \pi}{N L_1} T x_1}, \qquad T={\rm diag}( 1,1, \ldots, 1, 1-N) 
 \end{align}
 There exists a gauge transformation which allows one to switch from one gauge to the other \cite{GarciaPerez:2013idu, Poppitz:2022rxv}.  Which gauge to use is a matter of convenience. In dynamically abelianizing theories, $\Omega$ gauge is more convenient. To study the dynamics of the Yang-Mills theory on small symmetric $T^3$, $\Gamma$-gauge is better suited
\cite{GonzalezArroyo:1987ycm}. 
 However, the physical quantities such as vacuum energy density,  the value of 't Hooft flux $n_{12}$ mod $N$ are gauge invariant,  and we can use most  suitable gauge to calculate them.   In particular, whether $F_{12}$ is zero or not in 
 different regimes  on  $\R \times T^2 \times S^1 $ is a gauge invariant dynamical question, and different realizations occur.  See Fig.\ref{compete}.  

We now investigate the behaviour of the theory with unbroken $\Z_N^{[0]}$ symmetry on   $ \R^3  \times S^1_{\rm small}$  when it is compactified on  $ \R \times T^2 \times S^1_{\rm small}$ with $n_{12}=1$  flux.   
On $\R^3 \times S^1$, a center-stabilizing potential arises in many situations, such as QCD(adj) with  massless or light fermions  endowed with periodic boundary conditions \cite{Kovtun:2007py}, center-stabilizing double-trace deformation of Yang-Mills theory \cite{Unsal:2008ch}. The latter is equivalent  to  QCD(adj) with heavy adjoint fermions at  sufficiently small $L_3$ \cite{Cherman:2018mya,Unsal:2010qh}.   
We take the   deformation potential as: 
 \begin{align}
\Delta S(P_3) = - 2 \times S_{\rm 1-loop}^{\rm YM} 
  = 2 \times   \int_{M_4}   
     \sum_{n  \geq 1 }^{\lfloor  N/2 \rfloor} 
     \frac{2}{\pi^2 L_3^4}   \frac{1}{n^4} |\tr (P_3)^n|^2  \, 
\label{def-YM}     
\end{align}
The potential \eqref{def-YM} ensures that 
the minimum of $V_{\rm 1-loop}^{\rm YM}  + \Delta V$ is located 
at  
\begin{align} 
 P_3^{\rm sym.}& = \omega^{\alpha/2}  C_N 
   =   \omega^{\alpha/2}    
   {\rm Diag} \left(1,\omega,\omega^2, \ldots,\omega^{N-1} 
   \right),  \quad   
  \label{unbroken}
\end{align}   
where $\alpha=0, 1$ for $N$ odd/even, respectively.   

 We now consider the theory with  $n_{12}=1$ 't Hooft flux on  $ \R \times T^2_{\rm large} \times S^1_{\rm small}$.  
As mentioned above,  it is more convenient to use Abelian 
transition matrices.  Then, this leads to  
a non-flat field
strength  given by \cite{GarciaPerez:2013idu, Unsal:2020yeh,Cox:2021vsa,Poppitz:2022rxv}
\begin{align}
F_{12}^{(i)} =   \frac{2 \pi}{L_1 L_2} H^{(i)},  \qquad
H^{(i)} = {\rm Diag}(0, \ldots, 0, 1, 0, \ldots, 0) 
   - \frac{1}{N}{\bm 1}_N\, ,
\label{magflux}
\end{align} 
where $H^{(i)}$ is the $i^{\rm th}$ Cartan generator. Such a
configuration has a classical energy, given by 
\begin{align}
E_{\rm non-flat}^{\rm cl.} = \frac{1}{2 g^2}  \int_{T^2 \times S^1} 
    \tr\left[ \left(F_{12}^{(i)}\right)^2 \right]  
  =  \frac{L_3}{2  L_1L_2} \left(\frac{2 \pi}{g} \right)^2 
  \left(1 - \frac{1}{N} \right) \, . 
\label{vac-deg}
\end{align}
Yang-Mills theory with an abelian chromo-magnetic
background field has a well-known  tachyonic instability on $\R^4$, 
originally discovered by Nielsen-Olesen \cite{Nielsen:1978rm}. 
As a result of this consideration,  this configuration  was dismissed as a possible
ground state in Ref.~\cite{GarciaPerez:2013idu}.  However, we show below that this background is stable    and furthermore, corresponds to the ground state of the system on   $ \R \times T^2_{\rm large} \times S^1_{\rm small}$ \cite{Unsal:2020yeh}.

Let us first address the issue of stability.  For simplicity,  consider the $SU(2)$ gauge theory. 
For a magnetic background field $B= B_0 (\sigma_3/2)$, the mode decomposition of the field 
yields the frequencies  
\begin{align}
\omega_{k_z,n}^2 = \Big(  2B_0 
(n+ \half  +  S_3) +  k_z^2\Big)
\end{align}
 for the off-diagonal $W$-boson modes. The photons do not couple to external magnetic field. Clearly, $S_3=-1$, $n=0$ mode   is tachyonic
\cite{Nielsen:1978rm}. However, on $\R \times T^2 \times S^1$,
with $B_0 = \frac{2 \pi}{L_1 L_2}$ and with non-trivial holonomy 
in the $S^1$ direction, this formula is modified to 
\begin{align}
\omega_{k_3,n}^2 (\theta_{12})  = \Big(  2B_0 (n+ \half  
+  S_3) +  \frac{1}{L_3^2}  ( 2 \pi k_3 +   (\theta_{12} )^2   \Big) 
\label{NO}
\end{align}
 where in the center-symmetric holonomy
background  $\theta_{12} = \pi$.  Thus,  for sufficiently large $T^2$,  and with non-trivial holonomy,  
we have $\omega_{k_3,0}^2>0$  and 
  there is no  tachyonic instability. This is the assumption in the construction  made in \cite{Unsal:2020yeh}.
However, as $L_{T^2}$ gets smaller, at some point the 
tachyonic (Nielsen-Olesen) instability  will kick in. This 
is indicated by the red line in Fig.~\ref{fig:phases3d}. This 
is not a phase transition, as the theory is formulated
in small $\R \times T^3 $, and the low energy theory is essentially quantum mechanical.   But there is a drastic rearrangement 
of states. We do not explore this cross-over in detail in this 
paper. See \S.\ref{Sec:two-hierarchies} for comparison of the 
two-sides of the crossover.

Now, let us consider the opposite regime, starting with $\R^2 \times (T^2)_{\rm small}$ and compactify the theory further down to  $\R^1 \times (T^2)_{\rm small} \times S^1$. We assume $L_3 \gg L_{T^2}$.  In this regime, it is easier to work with non-abelian transition matrices,  and  the classical action is minimized for vanishing 
field strength  $F_{12} = 0$.  The Polyakov loops around the nontrivial 
cycles of $T^2$ are given  in terms of transition functiond and gauge holonomy.
\begin{align}
 P_1(x_2)&=g_1(x_2)\, 
    \mathcal{P}\rme^{\im \int_0^{L_1} a_1(x_1,x_2 )\diff x_1} 
    = S,\nonumber\\
 P_2(x_1)&=g_2(x_1)\, 
    \mathcal{P}\rme^{\im \int_{0}^{L_2}a_2(x_1,x_2)\diff x_2} 
    = C. 
\label{eq:unbroken_hol_con}
\end{align}
Since the classical field is gauge equivalent to $a=0$, the 
Polyakov loops are the same as the transition functions. The
flatness condition $(F_{12}=0)$ requires $P_1 P_3=P_3 P_1$ 
and $P_2 P_3=P_3 P_2$, and  hence, 
we have  
\begin{equation}
    P_3=\rme^{2\pi \im m/N}\bm{1},  \qquad  m=0,1,\ldots, N-1. 
\end{equation} 
For this configuration there is no classical energy, but the  
double trace deformation  \eqref{def-YM} costs  an energy 
that  is proportional to the the size of $T^2$. It  is  given by 
\begin{align}  
\Delta E_{\rm flat}  \propto  
     \frac{L_1L_2 }{L_3^2} \frac{1}{L_3}  N^2 .
\end{align}
This implies that for the two configurations on $T^3 = 
(T^2)_{12} \times (S^1)_3$ we obtain a classical energy 
cost for the non-flat gauge field and a one-loop energy 
cost for flat gauge field,
\begin{align}  
 \Delta E_{\rm non-flat}^{\rm cl.}  \propto \frac{N}{\lambda}   
  \left( \frac{L_3^2 }{L_1 L_2 }  \right) \frac{1}{L_3}, \qquad 
 \Delta E_{\rm  flat}^{\rm 1-loop}  \propto  
    N^2   \left( \frac{L_1L_2 }{L_3^2} \right)  \frac{1}{L_3}  ,
\label{costs} 
\end{align}
where $\lambda  \equiv g^2 N$ is 't Hooft coupling.  This leads to 
\begin{align}  
 \Delta E_{\rm non-flat}^{\rm cl.} \ll   \Delta E_{\rm  flat}^{\rm 1-loop}   \qquad L_{T^2} \gg L_3  \cr
  \Delta E_{\rm  flat}^{\rm 1-loop}   \ll     \Delta E_{\rm non-flat}^{\rm cl.}    \qquad L_3 \ll  L_{T^2}  
 \label{compare} 
\end{align}
Therefore, in the $n_{12}=1$ 't Hooft flux sector, the ground state for
 $L_{T^2} \gg L_3$  supports a non-zero  chromo-magnetic field,  $F_{12} \neq 0$, while 
    for  $L_{T^2} \ll L_3$,  the ground state is  a flat connection, $F_{12} =0$, as indicated in  Fig.\ref{compete}.

\noindent
{\bf Dynamics on $ \bm {\R\times T^2_{\rm large} \times S^1_{\rm small}}$:} In this regime, 
the theory is abelianized due to the non-trivial gauge 
holonomy \eqref{unbroken}. The dynamics at distances larger than
the inverse $W$-boson mass $(m_W)^{-1} = \frac{L_3N}{2 \pi}$ 
is described by an abelian gauge theory 
\begin{align}
SU(N) \rightarrow U(1)^{N-1} \qquad   
   P_3\; {\rm as\; adjoint \; Higgs \; field.}
\end{align} 
Despite the fact that $\tr (P_3^k)=0$ for $k \neq 0$ (mod $N$),
the zero form part of the center-symmetry $\Z_N^{[0]}$ on
\eqref{setup1-m} is \emph {broken} at the perturbative level. 
The order parameter that characterizes the broken symmetry 
is  $\tr ( P_3^k F_{12} )$, the Polyakov loop with an insertion 
of $F_{12}$. This operator acquires a non-vanishing expectation 
value 
\begin{align}
 \left( \frac{L_1 L_2}{2 \pi} \right) \;
\tr  \left( P_3 F_{12}^{(i)}\right)  = \omega^i,  
  \quad  i=1, \dots, N. 
\end{align} 
We denote these $N$ classical vacua by $| {\bm \nu}_i \rangle$, 
associated with the weights of the defining representation 
following \eqref{magflux}, and refer to them as \emph{magnetic
flux vacua} 
\begin{align} 
{\rm magnetic \;  flux \;  vacua}: 
 \quad \{ \; |  {\bm \nu}_i \rangle,  i=1, \dots, N\}\, . 
\end{align}
The theory admits monopole-instantons (which should be viewed 
as fractional instantons) interpolating between the perturbative
vacua
\begin{align}
| {\bm \nu}_{i+1}\rangle  \rightarrow 
          |{\bm \nu}_{i}\rangle, \qquad  \, . 
\end{align}
The change in the chromo-magnetic flux due to the tunneling
process is determined by the magnetic flux of a 
monopole-instanton.
\begin{align}
Q_{\rm mag}= {\bm \nu}_{i} - {\bm \nu}_{i+1} \equiv \alpha_i, 
   \qquad    i=1, \ldots, N.
\end{align}
The non-zero transition amplitudes 
\begin{align}
&    \langle {\bm \nu}_i | \rme^{-T\widehat{H}_{\mathrm{YM}}} 
   | {\bm \nu}_{i+1} \rangle 
      = K  \rme^{-S_I/N+ \im \theta/N}  
\label{tunneling}
\end{align}
between the perturbative vacua leads to the restoration of 
$(\Z_N^{[0]})_3$ center-symmetry non-perturbatively.   

 If the size of the torus is asymptotically large then we expect 
the dynamics on $\mathbb{R} \times(T^2)_{\rm large}\times 
(S^1)_{\rm small}$ to be the same as that of the theory on 
$\mathbb{R}^3 \times S^1$.  Let us briefly recall the EFT on $\mathbb{R}^3 \times S^1$ in the absence of 't Hooft flux \cite{Polyakov:1974ek,Unsal:2008ch}. The magnetic Coulomb gas 
is most 
easily described by using abelian duality, mapping $N-1$ photons 
into $N-1$ scalars $ * \diff{\bm \sigma} = L_3  \frac{2 \pi}
{ \im g^2} {\bm F}$. The monopole operator is given by 
\begin{align}
{\cal M}_i(x)= K \rme^{-S_m} 
  \rme^{\im \bm{\alpha}_i \cdot \bm{\sigma}(x)} 
\rme^{\im \theta/N}\quad (i=1,\ldots, N),  \quad S_m
  = \frac{S_I}{N}\, . 
\label{mon-op}
\end{align}
The  effective field theory of the the magnetic Coulomb gas 
is\footnote{\label{conventions} 
We use the following basis \cite{frappat:hal-00376660}. Let 
${\bm e}_i,  \; i=1, \ldots, N$  be  an $N$-component orthonormal 
basis in $\R^N$. Define weights of defining representation 
${\bm \nu}_i = {\bm e}_i - \frac{1}{N} \sum_{j=1}^N {\bm e}_j,\;
i=1, \ldots, N$. Then, ${\bm \nu}_i$ satisfy  ${\bm \nu}_i \cdot  
{\bm \nu}_j  = \delta_{ij} - 1/N$. We define  $N-1$ fundamental 
weights as   ${\bm \mu}_j  = \sum_{i=1}^{j}  {\bm \nu}_i, j=1, 
\ldots N-1$. These  ${\bm \mu}_j$ form a nice basis for the dual 
photons, ${\bm \sigma} = \sum_{i=1}^{N-1} \sigma_i \bm \mu_{i}$, 
because each $\sigma_i  \sim \sigma_i + 2 \pi$ is   $2 \pi$ 
periodic. In this basis $K_{ij} = {\bm \mu}_i \cdot {\bm \mu}_j 
= \min(i, j) - \frac{ij}{N} $ (called quadratic symmetric form),  
and in particular,  $ {\bm \mu}_i   \cdot {\bm \mu}_{N-1}=
\frac{i}{N}$. $K$ is the inverse of the Cartan matrix  $A_{ij} 
= {\bm \alpha}_i \cdot {\bm \alpha}_j = 2 \delta_{ij} -
\delta_{i,j\pm1} $,  $KA= \bm 1_{N-1}$.} 
\cite{Unsal:2008ch}
\begin{align}
 S_{\rm 3d}= \int_{\R^3} {g^2\over 8\pi^2 L_3} |\diff \bm{\sigma}|^2 
- 2 \zeta_m \sum_{i=1}^{N} 
  \cos\left(\bm{\alpha}_i \cdot \bm{\sigma}+{\theta \over N}\right).
\label{master}
\end{align}
This description is valid at energies below the $W$-boson mass, 
$m_W = \frac{2 \pi}{L_3N}$. The inverse of $m_W$ also controls 
the monopole size, $r_{\rm m} =  \frac{L_3N}{2 \pi}$. Note that 
the dual photon field $\bm{\sigma}$ has periodicity determined 
by the weight lattice,  $\bm{\sigma}\sim\bm{\sigma}+2\pi 
\bm{\mu}_i ,  \;  \bm{\mu}_i  \in \Gamma_w$ so that 
\begin{align}
\bm \sigma  \in \frac{\R^{N-1}}{2 \pi \Gamma_w}\, . 
\label{s-cell}
 \end{align}
The monopole gas generates a non-perturbative mass gap for the 
dual photon. For $\theta=0$ we have \cite{Unsal:2008ch}
\begin{align}  
m_p &=   \Lambda (\Lambda N L_3)^{5/6} \sin \frac{\pi p}{N},  
   \qquad  p=1, \ldots, N-1, \cr 
    \ell_D &= \frac{1}{m_{p=1}}  
    =  \frac{ N \ell_{YM}}{  (\Lambda L_3N)^{5/6} }
\label{Debye}
\end{align}  
where $\Lambda = \ell_{\rm YM}^{-1}$ is the strong scale 
of 4d Yang-Mills theory, and $\ell_D$ is the magnetic Debye
length. The small parameter $\epsilon \equiv (\Lambda N L_3)
\ll 1$ controls the semi-classical expansion on $\R^3 \times
S^1$.   

 We would like to understand the effect of the 't Hooft 
flux on the monopole EFT.  To this end, recall that Yang-Mills 
theory on $\R^3 \times S^1$ has a 1-form and 0-form symmetry
that descends from the 4d 1-form symmetry. 
\begin{align}
(\Z_N^{[1]})_{4d} \rightarrow 
     (\Z_N^{[1]})_{3d} \times  \Z_N^{[0]}\, . 
 \label{sym10}
 \end{align}
Turning on a two-form background field 
 \eqref{two-form}  associated with the 1-form symmetry
on $T^2$   modifies the 3d action by a topological
coupling as shown in \cite{Tanizaki:2019rbk}
\begin{align}
 S_{\rm 3d}= \int_{\R^3} {g^2\over 8\pi^2 L_3} 
   |\diff \bm{\sigma}|^2  
   - \int \frac{\im N}{2 \pi} \bm \mu_{N-1} \cdot 
     d  {\bm \sigma}   B 
 - 2 \zeta_m \sum_{i=1}^{N} 
  \cos\left(\bm{\alpha}_i \cdot \bm{\sigma}
     +{\theta \over N}\right).
\label{master-flux-2}
\end{align}
where $\bm\mu_{N-1} =\frac{1}{N}\left( 1,1,\ldots,1-N \right)$
is a fundamental weight. The topological coupling plays a
crucial role in the quantum mechanical limit.

\subsection{Separation of scales and transition from  3d to 1d EFT}

 It is useful to consider the physics of the effective field
theory on  $\R \times T^2_{\rm large} \times S^1_{\rm small}$
in two different regimes. First of all, because of the
compactification, the dual photon generates a Kaluza-Klein
tower, with separation $\frac{2 \pi}{L_{T^2}}$ between 
successive modes. The QFT regime is 
\begin{align}
 \frac{2 \pi}{L_{T^2}}  \ll m_{\rm gap}  \Rightarrow L_{T^2} 
 \gg  \ell_D  \qquad {\rm 3d \; EFT}.
\label{3deft} 
\end{align}
In this limit the modes are dense, and the dynamics is 
three-dimensional. In the opposite regime 
\begin{align}
 \frac{2 \pi}{L_{T^2}}  >  m_{\rm gap}  \Rightarrow L_{T^2} <  \ell_D  \qquad {\rm 1d \; EFT}, 
\label{1deft} 
\end{align}
only the zero mode of the dual photon field $\bm \sigma$
contributes. Note that the scale at which quantum mechanical 
1d effective field theory sets in is not a fixed scale on 
$\R \times T^2 \times S^1$. The critical value $L_{T^2}^*$
where this change occurs increases with decreasing $L_3$.   
We can describe the long-distance dynamics via a quantum
mechanical description at $L_{T^2} <  \frac{ N \ell_{YM}}
{(\Lambda L_3N)^{5/6} }$.

\begin{figure}[tbp] %
\begin{center}
\includegraphics[width= 0.7\textwidth]{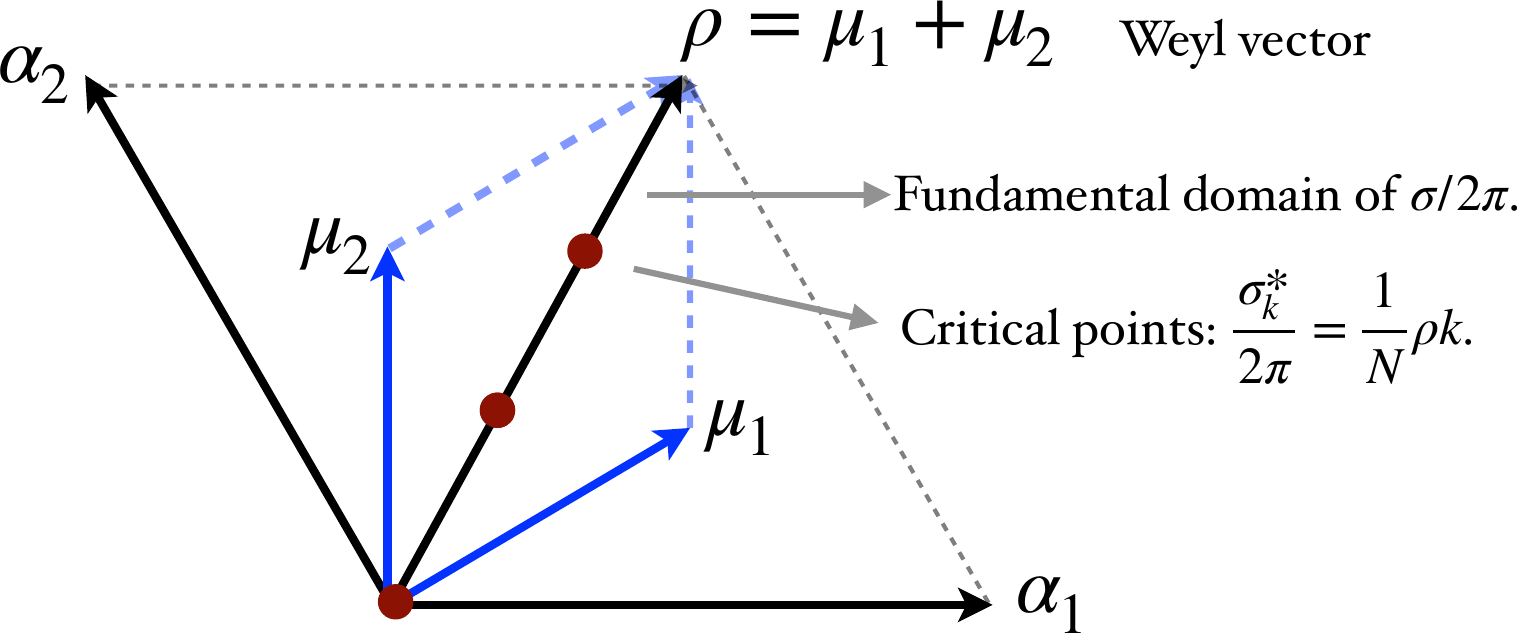} 
\end{center}
\caption{For $SU(3)$ gauge theory, the region bordered by 
the weight vectors $\mu_1, \mu_2$  is the fundamental cell 
of the weight lattice, and it is also the fundamental domain 
of ${\bm \sigma}/2 \pi$ \eqref{s-cell}. The red marked points
denote the meta-stable vacua of the $N=3$ theory  along the
Weyl vector $\rho$ according to \eqref{branch}.  These are 
the mean-fields (critical points) of the monopole-EFT and 
as one changes the $\theta$-angle, each one will become 
a genuine ground state at some value of $\theta$. We show 
that these are also the metastable vacua of the quantum
mechanical limit of the theory, providing adiabatic 
continuity between the two regimes. }
\label{fig:vacua}
\end{figure}

\noindent 
{\bf 3D EFT:}  As long as $T^2$ is sufficiently large so 
that \eqref{3deft} is satisfied one can use the EFT 
\eqref{master} in order to determine the dynamics. In
particular, the theta dependence of the vacuum energy 
density can be extracted from this description. For this
purpose we have to find the critical points of the potential 
in \eqref{master}, which are possible mean fields for the
${\bm \sigma}$ field. There are $N$ constant critical points
for the $\bm \sigma$ field in the fundamental domain 
\eqref{s-cell}. They can be found by using  the decomposition $\bm \sigma =
\sum_{i=1}^{N-1} \sigma_i {\bm \mu}_i $. The potential 
can then be written as 
\begin{align}
 V(\sigma_i)= -\zeta_m (e^{\im \sigma_1} + \ldots 
  +   e^{\im \sigma_{N-1}} 
  +  e^{-\im (\sigma_1  + \ldots+  \sigma_{N-1}  )} 
    + \rm {c.c})\, .  
\end{align}
The extrema are given by
\begin{align}
\frac{ \partial V} {\partial \sigma_i}= 0 
 \quad\Longrightarrow \quad 
 \sigma_i= 
    - (  \sigma_1 + \ldots + \sigma_{N-1})  \quad \forall i  
     \qquad \rm {mod} \; 2 \pi  \, . 
     \label{minpot}
\end{align}
Hence,  the $N$ critical points can be expressed as  
\begin{align}
\bm{\sigma}_k^*={2\pi\over N}k \bm{\rho} 
 \, \in \,  \frac{\R^{N-1}}{2 \pi \Gamma_w}\, , 
\label{branch}
\end{align}
where $ \bm{\rho}  =  \sum_{i=1}^{N-1}\bm{\mu}_i$ is 
the Weyl vector. These critical points are shown in
Fig.~\ref{fig:vacua} for $N=3$. Since $\bm \alpha_i. 
\bm \rho=1, i=1,\ldots, N$, the vacuum energy densities 
of these $N$ points  are given by 
\begin{align}
\mathcal{E}_k (\theta)=- 2 N  K \rme^{-S_I/N} 
 \cos\left({\theta+2\pi k \over N}\right). 
\label{fractional}
\end{align}
The true vacuum energy is given by $\mathcal{E}_{\rm ground}
(\theta) = \min_k \mathcal{E}_k (\theta)$, reflecting multi-branch structure and fractional theta angle dependence in Yang-Mills theory.  
 
 \subsection{Quantum mechanical limit as a deformed TQFT}
 \label{sec:QMlimit}

In the regime $L_{T^2} < \ell_D$ we can reduce \eqref{master-flux-2} 
to quantum mechanics. For this purpose, we have to integrate out all
modes except for the zero momentum mode in the Kaluza-Klein 
decomposition of the ${\bm \sigma}$ field.\footnote{
We would like to thank Yuya Tanizaki and Yui Hayashi for pointing out 
a flaw in this section in the first version of this manuscript.}  
In the present construction involving abelian gauge fields the 
twisted boundary conditions do not generate a classical mass for 
the dual photon, i.e. the classical moduli of sigma are not lifted 
by the $B$ field. Instead, they generate a new topological coupling.
In \S.\ref{sec:asym}, we will see an alternative construction in 
which the classical moduli is lifted and a  mass for dual photon is 
generated. Using  $\int_{T^2} B = \frac{2 \pi n_{12}}{N}$, we see 
that \eqref{master-flux-2} reduces to the following quantum 
mechanical action
\begin{align} 
 S_{\rm 1d}&= 
  S_{\rm kinetic} + S_{\rm top. coupling} + S_{\rm monopole}\, ,  \cr
  S_{\rm 1d}&= (L_1 L_2)  \int_{\R} {g^2\over 8\pi^2 L_3} 
       \left( \sum_{j=1}^{N-1} 
 K_{ij} \dot \sigma_i    \dot   \sigma_j  \right)
   -  \frac{\im  n_{12}}{2 \pi}   \sum_{j=1}^{N-1} \int 
         \frac{2 \pi j }{N}  \dot \sigma_j    \cr 
 & - 2 \zeta_m (L_1L_2)  \left( \sum_{j=1}^{N-1} 
  \cos\left( \sigma_j +{\theta \over N}\right) 
     +  \cos\left(  -(\sigma_1+ \ldots + \sigma_{N-1})  
    + {\theta \over N}\right)  \right)  \, , 
\label{master-new}
\end{align} 
where we used ${\bm \sigma} = \sum_{j=1}^{N-1} {\bm\mu_j} \sigma_j$.  
$K_{ij} = {\bm \mu}_i \cdot {\bm \mu}_j = \min(i,j) - \frac{ij}{N}$ 
is the quadratic symmetric form (the inverse of the Cartan matrix), 
and in the topological term, we used$^{\ref{conventions}}$
$ {\bm \mu}_i \cdot {\bm \mu}_{N-1}=\frac{i}{N}$.  This is the
dimensional reduction of the dual photon theory with a ``topological
coupling" induced by the background 2-form field $B$.\footnote{
The same quantum mechanical action also appears in two other QFTs. 
One is the compactification of $SU(N)$ level-$k$ WZW model on 
$\R \times S^1$ using twisted boundary conditions. The WZW term 
gives the topological coupling. The mass deformation in the WZW 
model correspond to monopole operators here. The other set-up is the
fermionization of this set-up. It is the $N$-flavor Schwinger 
model with a mass deformation \cite{Misumi:2019dwq}, see also
\cite{Tanizaki:2018xto,Kikuchi:2017pcp}.} 
Note that the $\theta$ angle appearing in the monopole operator is 
the theta angle of the 4d theory. In our set-up, apart from that,  
the topological coupling to the background $B$-field generates 
$N-1$ new theta angles, with values
\begin{align}
\theta_j= \frac{2 \pi j }{N}\, . 
\label{newtheta}
\end{align}
These new topological angles play a crucial role in what follows. In 
particular, let us first  treat  $S_{\rm monopole}$  as a small 
perturbation, which is justified in this regime. If we set it to 
zero and in the absence of $\theta_j$, the ${\bm \sigma}$ fields
describe the motion of a particle on a $T^{N-1}$ torus. In particular,
for $N=2$ it describes a free particle on a circle. The Hamiltonian
corresponding to $S_{\rm kinetic}$ in \eqref{master-new} can be
determined by a Legendre transform. We find 
\begin{align}
\frac{H}{C} =     \sum_{i, j=1}^{N-1}  A_{ij} p_i p_j  
  =  \sum_{j=1}^{N-2} (p_j - p_{j+1})^2 + p_1^2 + p_{N-1}^2 \, , 
\label{spec-Ham}
\end{align}
where $A_{ij} = {\bm \alpha}_i \cdot {\bm \alpha}_j = 2 \delta_{ij} 
-\delta_{i,j\pm1} $ is the Cartan matrix, $C=  \frac{L_3}{2 L_1L_2}
\left(\frac{2 \pi}{g} \right)^2$, and $p_i= \frac{1}{i} 
\frac{\partial}{\partial \sigma_i}$ are the conjugate momenta.   
In the absence of $\theta_j$, this system has a unique ground state.  

However, we can show that due to the additional theta angles in 
\eqref{newtheta} the non-interacting theory has an $N$-fold degenerate
ground state. $S_{\rm top. coupling}$ modifies \eqref{spec-Ham} by the
replacement $p_j \rightarrow p_j - \theta_j$: 
\begin{align}
   \frac{H}{C} =     \sum_{i, j=1}^{N-1}  A_{ij}  \left(p_i - n_{12}\frac{i}{N} \right) \left( p_j  - n_{12}\frac{j}{N} \right)
\end{align}
The energy spectrum of the free theory, $E_{\{m_n\}}$, is given 
in terms of $N-1$ quantum numbers,  $\{ m_n \} \in \Z^{N-1}$. 
Diagonalizing the Hamiltonian in the presence of $\theta_j$ we 
find  (for $N \geq 3$)
\begin{align}
\frac{E_{\{m_n\}}}{C} 
  =&     \sum_{j=1}^{N-2} \left[    
         \left(m_j - \frac{n_{12}\,j }{N}\right)    
      -  \left( m_{j+1} - \frac{n_{12}(j+1) }{N} \right)\right]^2 \cr
   &  + \left(  m_1  -  \frac{ n_{12}  }{N} \right)^2 
      +  \left(  m_{N-1}  -  \frac{n_{12} (N-1)}{N} \right)^2 
        \, .
\label{spec}
\end{align} 
For $n_{12}=1$, the ground state is $N$-fold degenerate and the 
vacua are given by
\begin{align}
|m_1, \ldots, m_{N-1} \rangle = \{  |0, \ldots, 0 \rangle,  
  |0, \ldots, 1 \rangle,  
  |0, \ldots, 1,1 \rangle,  \ldots   
  |1, \ldots, 1 \rangle,    \} \, . 
\label{spec-zero}
\end{align}
These ground state energies coincide with \eqref{vac-deg}, 
as expected.

These vacua are just generalization of the free particle on a 
circle problem with theta angle $\theta_1=\pi$.  In fact, for  
$N=2$, we have
\begin{align}
 E_{m_1} &=  {\frac{C}{2}}  \left( m_1- \frac{1}{2} \right)^2  \, .   
\label{spec2}
\end{align}
The vacua  are two-fold degenerate and given by $m_1=0,1$. In a 
coordinate basis, denoting the position of the particle by $\sigma$, 
the wave functions are $ \langle  \sigma |0\rangle 
= {1 \over \sqrt{2 \pi} }$ and $ \langle 0|1\rangle 
= { \rme^{\im \sigma} \over \sqrt{2 \pi} }$.\footnote{
It is interesting to note that the monopole operators correspond 
to the wave function of a particle on a torus, $T^{N-1}$.} 
In the general case the coordinate space $(\sigma_1, \ldots, 
\sigma_{N-1})$ expressions for the states in the Hilbert space 
are given by
\begin{align}
 \langle \sigma_1, \ldots, \sigma_{N-1} |m_1, \ldots, m_{N-1} \rangle  
 =  \rme^{\im m_1 \sigma_1}  \ldots \rme^{\im m_{N-1} \sigma_{N-1} }.
\label{spec-coor}
\end{align}
The coordinate representation also clarifies the role of the monopole
operators. Monopole operators act as
\begin{align}
 \langle \tilde m_1, \ldots, \tilde m_{N-1}  |  
    \rme^{\im \sigma_n}  |m_1, \ldots, m_{N-1} \rangle  
    =   \delta_{  \tilde m_1,  m_1} \ldots 
        \delta_{  \tilde m_n,  m_n +1} \ldots   
        \delta_{  \tilde m_{N-1},  m_{N-1}}\, . 
\label{spec-mon}
\end{align}
We can identify these states with the magnetic flux states described 
on an intuitive basis in \cite{Unsal:2020yeh}. The identification and 
the role of the monopole are
\begin{align}
 \underbrace{ |0, \ldots, 0 \rangle }_{ |{\bm \nu}_N  \rangle}    
 \xrightarrow[  ] {  \rme^{\im \sigma_{N-1}}  }   
 \underbrace{ |0, \ldots, 1 \rangle}_{ |{\bm \nu}_{N-1}  \rangle}     
 \xrightarrow[  ] {  \rme^{\im \sigma_{N-2}}  }  
 \underbrace{ |0, \ldots, 1,1 \rangle }_{ |{\bm \nu}_{N-2}  \rangle}   
     \ldots      
 \xrightarrow[  ] {  \rme^{\im \sigma_1}   }   
 \underbrace{|1, \ldots, 1 \rangle}_{ |{\bm \nu}_1  \rangle}     
 \xrightarrow[  ] {  \rme^{\im \sigma_N}  }  
\underbrace{ |0, \ldots, 0 \rangle   }_{ |{\bm \nu}_N  \rangle}  \, . 
\label{states}
\end{align}
This shows that the $N-1$ topological couplings in the quantum 
mechanical model act as  $N-1$  charges  (like $\pm {\bm \nu}_1)$ 
at infinity. In this way, tunneling happens between $N$ perturbative
ground states that share the same energy.\footnote{
Consider a parallel plate capacitor with charges $Q = \pm 1$. The 
energy density is non-zero inside the capacitor and zero outside.
Insert a pair of charges at $\pm \infty$, respectively, $\mp \half$.
Now, the electric energy density inside the capacitor and outside 
are the same.  It is well-known in the context of the Schwinger
model that theta angle can have a physical interpretation as charge 
at infinity \cite{Coleman:1976uz}.  The topological couplings  
$\tilde \theta_j$ or equivalently,  the background magnetic flux 
${\bm \nu}_i$ play the same role here.}

\noindent
{\bf 1D EFT, Hamiltonian in terms of magnetic (or electric) flux 
basis:} The simplest way to determine the spectrum in the quantum
mechanical limit is to use the tight-binding approximation. 
The representation \eqref{states} makes it manifest that we can 
represent the monopole operators on the space of perturbative 
vacua as 
\begin{align}
\rme^{\im \sigma_i} + {\rm h.c.}  \;  :  \; 
  |{\bm \nu}_i \rangle \langle {\bm \nu}_{i+1}  |  + {\rm h.c.} 
\end{align}
The problem reduces (within the Born-Oppenheimer approximation) to
a particle on a circle with $N$ degenerate harmonic minima. The 
tight-binding Hamiltonian is
\begin{align}
 H = \sum_{i=1}^{N} E_0 |{\bm \nu}_i\rangle \langle {\bm\nu}_{i} |  
   -  \sum_{i=1}^{N}  K \rme^{-S_m - \im \theta/N} \  
    |{\bm \nu}_i \rangle \langle {\bm \nu}_{i+1}|  
      + {\rm h.c.} \, , 
\label{Hamiltonian}
\end{align} 
where the second term describes nearest neighbors hopping 
between adjacent flux vacua \eqref{states}.  The first 
term is the zero point energy $E_0$, given by $E_{\rm non-flat}
^{\rm cl.}$ in \eqref{vac-deg}. We can diagonalize the Hamiltonian
\eqref{Hamiltonian} via a discrete Fourier transform: 
\begin{align}
 | \widetilde k \rangle = \frac{1}{\sqrt N} 
      \sum_{i=1}^{N} \omega^{ki} | {\bm \nu}_i\rangle
\end{align}
and obtain 
\begin{align}
    H= \sum_{k=0}^{N-1} {\varepsilon}_k (\theta) 
    | \widetilde k \rangle \langle  \widetilde k |
\end{align}
where ${\varepsilon}_k (\theta)$ is given in \eqref{fractional}.  
We can interpret the states $| \widetilde k \rangle$ as the
{\it electric flux states}. The low energy limit of the 
quantum system can be described either in terms of the 
magnetic or electric basis
\begin{align}
 {\rm Magnetic \;flux \;  states\; }: \; \quad 
  | {\bm \nu}_i \rangle  
  \quad \Longleftrightarrow
  \quad  | \widetilde k \rangle \, : \quad 
    {\rm Electric \;flux \;  states\; }  
\end{align}
The string tension for charge $q$ in the the regime $- \pi < 
\theta <\pi$ is given by $T_q= \varepsilon_q(\theta) -
\varepsilon_q(0)$. The Born-Oppenheimer appropximation is 
justified because perturbative gaps are parametrically larger 
than the non-perturbative ones.

We observe that the vacuum structure of the theory on  
$L_{T^2} \gg  \ell_D $ is continuously connected to the one 
on  $L_{T^2} \ll  \ell_D $, and monopoles continue to 
determine non-perturbative dynamics even in the quantum 
mechanical regime.

\subsection{Monopole interaction in the QM limit}

  It is interesting to note that the monopole operator is unchanged
in the QM limit. The interaction between monopoles on $\R \times 
T^2_{\rm large} \times S^1_{\rm small}$ or in the $\R^3 \times S^1$
limit is the $1/r$ Coulomb interaction in 3d.  Because the photons
remain ungapped classically, the interaction of the monopoles in 
the quantum mechanical limit is  dictated by the Coulomb
interaction in 1d. Considering the correlation functions in 
the 3d and 1d limit,
\begin{align} 
\langle {\cal M}_i( {\bm x}) {\cal M}_j( {\bm 0}) 
    \rangle_{\rm free, \; 3d} 
 & =  \langle \rme^{\im \bm{\alpha}_i \cdot \bm{\sigma}({\bm x})} 
   \rme^{\im \bm{\alpha}_j \cdot \bm{\sigma}({\bm 0})} \rangle   \cr
\langle {\cal M}_i(\tau) {\cal M}_j(0) 
       \rangle_{\rm free, \;  1d} 
 & =  \langle \rme^{\im \bm{\alpha}_i \cdot \bm{\sigma}(\tau)}
   \rme^{\im \bm{\alpha}_j \cdot \bm{\sigma}(0)} \rangle \, ,  
\label{int1}
\end{align}
we find that the interactions between monopoles in the 3d and 1d 
EFTs\footnote{
The formula in 1d is valid at distances $\tau \gg L_{T^2}/2\pi$.  
To get a formula which interpolates between the 3d interaction 
and 1d interaction, we have to incorporate Kaluza-Klein 
modes. See also the footnote underneath \eqref{int-exp}. }
are given by 
\begin{align} 
V({\bm x}) &= A_{ij} \left( \frac{1}{4 \pi |{\bm x}|} \right) \cr
V(\tau)  &  
 = \frac{ A_{ij} }{(L_1 L_2)} \left(- \frac{1}{2} |\tau|   \right)
\label{int2}
\end{align}
where the pre-factors 
\begin{align} 
 A_{ij}= L_3  
   \left( \frac{2 \pi}{g} \right)^2 \alpha_i \cdot \alpha_j   \, ,
\label{int2-coef}
\end{align}
arises from the charges of the monopoles  and the normalization 
of the kinetic term. 

  The monopole-monopole interaction in the quantum mechanical 
limit remains  long-ranged to all orders in perturbation theory. 
The system is a multi-component Coulomb gas in $d=1$. The effect 
of the monopoles and bions, as it is the case on $\R^3 \times S^1$
is to render the dual photon massive, via a magnetic Debye 
mechanism in 1d.  If we insert an external magnetic charge 
into the system, its field will decay  exponentially as  
$ \frac{1}{2 \kappa} \rme^{-\kappa |\tau| } $ where $\kappa \sim 
\rme^{-S_m/2}$ due to Debye screening.  In fact, the natural 
frequencies of the quantum mechanical system $\omega_p, =1, 
\ldots, N-1$  are exactly the mass spectrum of the theory 
\eqref{Debye} on $\R^3 \times S^1$, $\omega_p=m_p$. This shows 
that the most crucial property of the magnetic Coulomb gas 
remains the same in the reduction from $d=3$ to $d=1$.

\section{Semi-classics on $\R \times T^2_{\rm small} \times 
S^1_{\rm large} $ with 't Hooft flux through $T^2$}
\label{sec:T2small}

 In this section we consider compactifactions on a small
torus, corresponding to the gray band along the $x$-axis in 
Fig.~\ref{fig:phases3d}. In \eqref{costs} we showed that 
the energy costs of 't Hooft flux $n_{12}=1$ sector 
with a  non-flat field strength is proportional to $\Delta 
E_{\rm non-flat} \sim (N/ \lambda)(L_3/L_{T^2})^2$, while 
the cost of realizing it with a flat field strength is 
proportional to $\Delta E_{\rm flat} = N^2 (L_{T^2}/L_3)^2$.
Therefore, if  $L_{T^2} \lesssim \frac{ L_3} {N^{1/4} 
\lambda^{1/4}}$,  the vacuum of the theory is described by flat gauge fields. 
  This crossover may well be related to the
Nielsen-Olesen instability as mentioned around \eqref{NO}, 
but we postpone a detailed study of this issue to future work.

Assume that $L_3 \rightarrow \infty$  is very large.  Then
the setting is
\begin{align} 
M_4 =  \lim_{L_3 \rightarrow \infty}  
   \mathbb{R}\times \underbrace{T^2}_{\rm short}
   \times \underbrace{(S^1)_3}_{\rm large} 
   =  \R^2 \times T^2
   \, .
\label{setup3}
\end{align}
The short symmetric torus has size $L_1=L_2 \ll \Lambda^{-1}$  
and the large $(S^1)_3$ has size $L_3 \gg L_1$.  The semi-classical 
analysis below is an overview of necessary ingredients 
from \cite{Tanizaki:2022ngt}.  

\subsection{Overview of $\R^2 \times T^2_{\rm small}$: 2d EFT is a
deformed TQFT} 

Consider $n_{12}=1$ unit of 't~Hooft flux along $T^2$. Based 
on the general analysis of \cite{tHooft:1979rtg}, we can choose 
the transition functions to be $g_1=S$ and $g_2= C$ shift and 
clock matrices. By the relation between Polyakov loop and 
transition amplitude \eqref{eq:unbroken_hol_con}, this implies 
that the gauge holonomies (Polyakov loops) in the $1,2$ 
directions are given by
\begin{align} 
P_1=S,   \qquad P_2=C.  
\label{CS}
\end{align} 
Therefore, the theory on $\R^2 \times T^2 $ may be viewed as a 
2d Yang-Mills theory coupled to two adjoint Higgs fields, $P_1$ 
and $P_2$ \cite{Tanizaki:2022ngt}. In fact, from the twisted 
Eguchi-Kawai reduction, we know that the twisted boundary conditions 
can be replaced by periodic boundary conditions, but with a modified 
gauge action \cite{GonzalezArroyo:1982ub, GonzalezArroyo:1982hz}. 
In our context, for small $\R^2 \times T^2 $,  this translates to 
the  ``classical potential" 
\begin{align}
\tr |P_1 P_2 -  \omega P_2 P_1|^2, \qquad 
\omega= e^{i \frac{2 \pi}{N}}\, . 
\end{align}
The perturbative analysis implies that, in analogy to the 
charge $N$ abelian Higgs model in which the $U(1)$ gauge group 
is reduced to $\Z_N$ by a Higgs mechanism, the long distance 
 gauge structure of the theory on small $\R^2 \times T^2 $ is reduced to $\Z_N$:
\begin{align} 
SU(N) \xrightarrow{P_1, P_2\, \text{adjoint Higgs}}\Z_N \, .
\label{Higgsing}
\end{align}
One can think of this result as an adjoint Higgs regime induced 
by $P_1$ which reduces the $SU(N)$ gauge structure to $U(1)^{N-1}$, 
followed by a Higgs mechanism triggered by $P_2$ which further 
reduces $U(1)^{N-1}$ down to $\Z_N$, 
\begin{align}
SU(N) \xrightarrow{P_1\, \text{adjoint Higgs}}  U(1)^{N-1}  
      \xrightarrow{P_2\, \text{adjoint Higgs}} \Z_N 
\label{2-stage}
\end{align} 
However, since we are using a symmetric torus, the two Higgs 
mechanisms occur at the same energy (length) scale. There is no
advantage gained here by viewing the adjoint Higgs mechanism as a 
two-step process.  In the next section, we will consider 
an asymmetric torus and a two-step Higgs mechanism. 
 
  Because of  \eqref{Higgsing}, the 
long distance theory may be described by a 2D gauge theory with 
gauge group $\Z_N$. Moreover, the twisted boundary conditions 
force all the gluons to acquire tree-level masses. We quickly 
recall how this happens since we will use the result in the next
section too.  We can express the gauge fields as      
\begin{align}  
 a =  \sum   a^{(\bm{p})} J_{\bm{p}}. \quad 
 J_{\bm{p}} = \omega^{-p_1 p_2/2} C^{-p_1} S^{p_2}, \qquad 
 \bm{p} =(p_1,p_2)\in (\mathbb{Z}_N)^2 \setminus \{\bf 0\}\, , 
\label{basis1}
\end{align} 
where  $J_{\bm{p}}$ is a suitable Lie algebra basis, similar 
to what is used in the perturbative analysis of the twisted 
Eguchi-Kawai model \cite{GonzalezArroyo:1982hz}. Denoting the 
Fourier components of $a^{(\bm{p})}$ by $a^{(\bm{p, k})}$, 
where  $\bm{k} \in (\mathbb{Z})^2$, we can show that all 
modes acquire non-zero masses given by 
\begin{equation}
M_{\bm{p},\bm{k}}^2=\left({2\pi\over NL_{T^2}}\right)^2
  \Bigl((Nk_1+p_1)^2+(Nk_2+p_2)^2\Bigr). 
\label{gluonspectrum}
\end{equation}
Therefore, gluons do not have any zero modes, and the lowest 
gluon mass is $2\pi/NL_{T^2}$ \cite{GonzalezArroyo:1982hz}. 
The spectrum is sketched in  Fig.~\ref{fig:stage} (left).  

 Recall that in the $\R^2 \times T^2$ compactification, the 
1-form symmetry of the 4d theory decompose into 1-form and 
0-form symmetries in 2d as
\begin{align}
(\Z_N^{[1]})_{4d} \xrightarrow[]{ {\rm on}\; {\R^2 \times T^2}} 
  (\Z_N^{[1]})_{2d} \times  \Z_N^{[0]} \times  (\Z_N^{[0]})\, . 
\end{align}
Although the 0-form symmetries remain unbroken to all orders in
perturbation theory, the 1-form part of the center symmetry is 
spontaneously broken, and the effective 2d theory is perturbatively 
a {\bf topological  $\Z_N$ gauge theory}. The main question is 
whether this perturbative conclusion  is stable 
or unstable non-perturbatively. 

Indeed, the proliferation of center-vortices leads to linear 
confinement. The center-vortices in 2d proliferate because they 
are fractional instantons with finite action
$S_{\rm I}/N$ \cite{Gonzalez-Arroyo:1998hjb, Montero:1999by, 
Montero:2000pb}.  Their density in the vacuum of the theory 
on $\R^2$ is proportional to $e^{-S_I/N}$.  Furthermore, center 
vortices  have non-trivial mutual statistics with the Wilson 
loops. Let ${\bm x} \in \R^2$ be the location of the center of the 
center-vortex, and $C$ be the boundary of a disk $D$. Then, a
Wilson loop $W_{\cal{R}}(C)$ in the $SU(N)$ representation 
$\cal{R}$ obeys 
\begin{equation}
W_{\cal R}(C) =
\begin{cases}
    \omega^{k} & \text{if}\quad {\bm x} \in D, \\
      1        & \text{if}\quad {\bm x} \notin D, 
\end{cases} 
\label{VW}
\end{equation}
where $k$ is the $N$-ality of ${\cal R}$. 
 
The counterpart of the EFT in 2d for the center-vortex theory is 
a topological BF-theory deformed by local topological operators  
\cite{Nguyen:2024ikq,Cherman:2021nox}.  
\begin{align}
\label{TQFT1}
S_{\rm 2d}=  & S_* + \Delta S  
  =    \frac{iN}{2 \pi} \int_M \varphi \, da  
    - 2\zeta_{\rm v} \int_M d^2 x \, \cos(\varphi + \theta/N) \, .
\end{align}
Here, $\varphi$ is a compact scalar field ($\varphi \sim \varphi 
+ 2 \pi$) and $a$ is a $U(1)$ gauge field. The deformation 
$\Delta S$ describes the proliferation of center-vortices. 
The parameter $\zeta_{\rm v} \sim  e^{-S_{\rm I}/N }$
is the density of the center-vortex gas. 

The partition function localizes on $N$ critical points, 
and is given by 
\begin{align}
Z(\theta) = \int D\varphi Da \;  e^{-S} 
  =   \sum_{k=0}^{N-1} e^{- \varepsilon_k {\cal A}_{M_2} }, \qquad  
  \varepsilon_k = -2 \zeta_{\rm v} 
       \cos\left( \frac{2 \pi k +\theta}{N} \right)\, . 
  \label{DTQFT}
\end{align}
The non-degeneracy of the $N$ energy levels implies confinement. 
To see this, we need to evaluate the expectation value of the 
Wilson loop.  For $\theta \in (-\pi,\pi]$, 
\begin{equation}
 \langle W^q(C) \rangle \sim 
  \exp \{- (\varepsilon_{q} - \varepsilon_{0}) {\cal A} \}.
    \label{eq:AreaLaw}
\end{equation}
where  $\mathcal{A}$ is the area enclosed by $C$.  
For $\theta \in (-\pi,\pi]$,  
the string tension associated
with the charge $q$ Wilson loop is $T_q = \varepsilon_q - 
\varepsilon_0 $.

This construction provides a semi-classical realization 
\cite{Tanizaki:2022ngt} of the  center vortex  mechanism 
\cite{DelDebbio:1996lih,DelDebbio:1998luz,Engelhardt:1998wu,
Alexandrou:1999iy,Alexandrou:1999vx,deForcrand:1999our,
deForcrand:2000pg,Sale:2022qfn} in a 2d EFT. However, we also 
note that in $d \geq 3$, the proliferation of center-vortices 
is a necessary, but not a sufficient condition for confinement  
as shown in \cite{Nguyen:2024ikq}, while proliferation of 
monopoles is sufficient.

\subsection{$\R \times T^2_{\rm small} \times S^1$ } 

 So far, we considered $\R^2 \times T^2_{\rm small}$. Now, we 
compactify an extra circle, and consider the compactification of 
the theory on  $\R \times T^2_{\rm small} \times (S^1)_3 $.  We
assume 
\begin{align} 
 L_{T^2} =L_1=L_2   \ll  L_{3} 
\end{align} 
With this assumption, we can work with flat gauge fields to 
explore low energy properties. Let us first describe the classical
vacua. As before, we denote the holonomy of the gauge field along
long $(S^1)_3$ as $P_3$. The classical vacua are described by
configurations of $P_3$ that commute with $P_1=S$ and $P_2=C$. 
Since $C$ and $S$ can be used to generate the Lie algebra of 
$\mathfrak{su}(N)$,  $P_3$ must be proportional to the identity
matrix $\rme^{\im \alpha} \bm{1}$. Since $\det P_3=1$, possible 
values of $ P_3=\rme^{2\pi \im m/N}\bm{1}$  ($m=0,1,\ldots, 
N-1$), 
\begin{align}
 [P_3, C] = [P_3, S] =0,  \qquad   
 \det  P_3=1  \qquad  
 \Longrightarrow   \qquad P_3=\rme^{2\pi \im m/N}\bm{1} \, .
\end{align} 
Let us denote these classical vacua by $|m\rangle$, obeying  
\begin{align}
 P_3 |m\rangle =\rme^{2\pi \im m/N} |m\rangle\, . 
\end{align} 
Therefore, perturbatively, the $(\Z_N^{[0]})_3$ center symmetry 
is spontaneously broken.  
 
 If we ignore the possibility of tunneling, we obtain a quantum
mechanical system with $N$ vacua. The Hamiltonian is just a 
constant, which we can set to zero. Tunneling events lift the
degeneracy of the vacua, similar to  \eqref{Hamiltonian}
\begin{align}
   H = \sum_{m=1}^{N} \tilde E_0 | m \rangle \langle m |     
    -  \sum_{m=1}^{N}  K \rme^{-S_{\rm v} 
    - \im \theta/N} | m \rangle \langle m+1 |  + {\rm h.c.} \, ,
\label{Hamiltonian1}
\end{align} 
where  the zero point energy $\tilde E_0$ is the 
$E_{\rm flat}^{\rm 1-loop}$ in \eqref{costs}. The effective 
action in this regime is the compactification of the deformed 
TQFT  \eqref{TQFT1}, and is given by
\begin{align}
\label{TQFTvortex-red}
S_{\rm 1d} =  &  \frac{iN}{2 \pi} 
   \int dt  \varphi \, \frac{d\alpha}{dt}  
   - 2(\zeta_{\rm v} L_3N) \cos(\varphi + \theta/N) \, . 
\end{align}
Note, however, that the states entering \eqref{Hamiltonian1} 
and \eqref{Hamiltonian} are quite different. In the next 
subsection,  we compare the two regimes leading to QM.

\subsection{Two hierarchies leading to QM, dramatic reordering 
(same EFT)} 
\label{Sec:two-hierarchies}

 In the previous sections we considered two regimes in which the
 theory on small  $\R \times T^3$ with $n_{12}=1$ reduces to quantum
mechanics.  We studied $\R \times T^2 \times S^1$, with both 
$T^2$ and $S^1$ small (smaller than the Yang-Mills length scale 
$\Lambda^{-1}$), but with opposite hierarchies of their sizes, 
$ L_{T^2} \ll L_{S^1}$ vs. $L_{S^1} \ll L_{T^2}$. Below we list 
the physical properties we obtained in these limits, including 
the Polyakov loop $P_3$, the field strength $F_{12}$, the 
perturbative vacua, tunnelings events, tunneling amplitudes, 
and the electric flux states and energies. The mapping between 
these two semi-classical domains was also examined in
\cite{Poppitz:2022rxv}, and the matching of mixed anomalies 
was studied in \cite{Tanizaki:2022ngt,Cox:2021vsa}. Our analysis
demonstrates that both QM regimes are described by a $\Z_N$
topological theory deformed by a fractional instanton operator. 
The summary of these two regimes is shown in the following tables. 
\begin{center}
\hspace{-.0cm}
\begin{tabular}{||l|l |l|l|l|l|} 
 \hline
 Regime & Mechanism  & Holonomy & Field strength & Pert. vacua &
    States \\  
 \hline
   \;  $L_{T^2} \ll L_{S^1}  $ & vortices
    & $ \tr P_3  \propto  e^{\im  \frac{2 \pi}{N} m}$ 
    & $F_{12}=0$  
    & $ \tr  ( P_3 )  \propto  e^{\im \frac{2 \pi}{N} m} $ 
    &  $  | m \rangle$     \\ 
\;   $L_{S^1} \ll L_{T^2}  $  & monopoles
    & $\tr P_3 = 0 $ 
    & $F_{12}=   \frac{2 \pi H^{(i)} }{g L_1 L_2}  $  
    & $ \tr  ( P_3 F_{12}) \propto  e^{\im \frac{2 \pi}{N} i} $ 
    &  $ |{\bm \nu}_i \rangle $  \\ 
 \hline
\end{tabular}
\end{center}
\begin{center}
\hspace{1.0cm}
\begin{tabular}{||l|l|l |l |l|} 
 \hline
 Tunnelings   &  Tunneling amp. &  vevs & E-flux states  
    & Energies \\ 
 \hline
 $| m \rangle \rightarrow   | m+1  \rangle$       
    &  ${\rm Exp}[-8 \pi^2/(g^2N)]$   & $\langle\tr  ( P_3 )   
      \rangle$=0
    &  $| \widetilde q \rangle\propto \sum_{m} \omega^{qm} |m  
      \rangle$
    &  $ \widetilde E_0 +  \varepsilon_k (\theta)$    \\ 
 $|{\bm \nu}_i \rangle   \rightarrow | {\bm \nu}_{i+1}  \rangle $   
    &  ${\rm Exp}[-8 \pi^2/(g^2N)]$   & $\langle\tr  ( P_3 F_{12} ) 
     \rangle$=0
    &  $|\widetilde q \rangle\propto\sum_{i} \omega^{qi} | {\bm \nu}
    _i\rangle$  
    &  $ E_0 +  \varepsilon_k (\theta)  $   \\ 
 \hline
\end{tabular}
\end{center}
The zero point energy $\widetilde E_0$ 
is equal to $E_{\rm flat}^{\rm 1-loop}$ and $E_0$ is given by
$E_{\rm flat}^{\rm cl.}$ in \eqref{costs}. 


Despite the dramatic change of the Polyakov loop and the field 
strengths as $M_4$ changes from $ L_{S^1} \ll L_{T^2}$ to 
$L_{T^2} \ll L_{S^1}$ there is no  phase transition, but a  
reordering of vacua in quantum mechanics. As explained around \eqref{NO}, 
 we believe that this dramatic reordering 
is related to Nielsen-Olesen instability \cite{Nielsen:1978rm}, 
which in turn must be related to which one of the options in 
\eqref{costs} is the lower energy configuration. For example, 
for SU(2) gauge theory, the frequencies in mode decomposition 
of gluon field with spin $S_3 =\pm 1$ are \cite{Poppitz:2022rxv}
\begin{align}
\omega_{k_3,n, S_3}^2 (\theta_{12})  = 
  \Big(  \frac{4 \pi}{L_1 L_2} (n+ {\half}  +  S_3) 
   +  \frac{1}{L_3^2}  ( 2 \pi k_3 +   (\theta_{12}) )^2   \Big), 
   \qquad  n\in  {\mathbb N}^0, 
 \; k_3 \in \Z
\end{align}
which has a mode that can become tachyonic $(n=0, S_3=-1)$.   
In the regime $L_3 \ll L_{T^2}$ with non-trivial holonomy 
$\theta_{12}=\pi$,  $\omega_{k_3,n, S_3}^2 >0$, the would-be
tachyonic mode is lifted, and the $F_{12} \neq 0$ background 
is stable ground state. In the $ L_{T^2} \ll L_{S^1}$ regime,  for $F_{12} 
\neq 0$, $\omega_{k_3, n=0, S_3=-1}^2 (\theta_{12}) 
<0$, and there is an instability. The ground state has vanishing 
chromomagnetic field. 

 The cross-over between the two description is shown as 
the red dashed line in  Fig.\ref{fig:phases3d}. The nature 
of the topological defects in these two regimes are different. 
In the $|m\rangle$ basis, there are tunneling events between 
trivial holonomy vacua $\tr (P_3)\propto e^{\im\frac{2\pi}{N} m}$. 
Only at the core of the center-vortex, the holonomy becomes 
non-trivial. In the $|{\bm \nu}_i\rangle$ basis there are 
tunneling events between nontrivial holonomy vacua. At the 
core of the center-vortex two eigenvalues of the holonomy 
becomes degenerate. 

 Both quantum mechanical limits are described in the deep 
infrared by a deformed $\Z_N$ TQFT in 1d. This is consistent 
with the idea that in the diagram Fig.~\ref{fig:phases3d}
the $\R^2 \times T^2$ regime is adiabatically connected 
to the $\R^3 \times S^1$ regime.

\section{Two-scale adjoint Higgs mechanism and monopole-vortex 
continuity: From 3d to 2d}
\label{sec:asym}

In this section, we consider a different interpolation between
the semi-classical regimes described by monopoles and vortices. 
Following Hayashi and Tanizaki we consider $\R^2 \times (S^1)_2 
\times (S^1)_3$ and study the theory as the size of the circle
$(S^1)_2$ is varied \cite{Hayashi:2024yjc} at fixed small $(S^1)_3$. The main new 
result in this section is a derivation of the 2d  EFT 
\eqref{QFT-master} in terms of dual photon field. 
   
\subsection{Connecting the monopole and vortex EFTs}
   
 It is useful to construct an adjoint Higgs mechanism that
generates a {\bf parametric scale separation} between the scale 
for abelianization  $SU(N) \rightarrow U(1)^{N-1}$ and the scale 
at which  $U(1)^{N-1} \rightarrow \Z_N$. In this case we can use 
the fields of  effective field theory based on the $U(1)^{N-1}$
gauge structure  to reach the $\Z_N$ gauge theory.

 In the infinite volume field theory on $\R^4$, this is rather
standard. Consider two adjoint scalars $(\Phi_1, \Phi_2)$ with a 
potential that favors a non-commuting minimum, such as the clock 
and shift matrices, and arrange a scale separation. This will 
lead to an abelian theory at some high energy scale $E_{1}$ 
and $\Z_N$ gauge theory at some lower energy scale $E_2$.  
\begin{align}
SU(N) \xrightarrow [ {\rm scale} \; E_1 ] 
      {\Phi_1\,\text{adjoint Higgs}}\; U(1)^{N-1} 
      \xrightarrow [ {\rm scale} \; E_2 ] 
      {\Phi_2\,\text{adjoint Higgs}} \; \Z_N, 
      \qquad E_2 \ll E_1
\end{align} 
The physics on $\R^d$ at low-energies $E < E_2$ is described
by $\Z_N$ gauge theory, intermediate scales $E_2 < E < E_1$ are
governed by $U(1)^{N-1}$ theory, and high energy $E> E_1$ can be
described in terms of the original $SU(N)$ degrees of freedom. 
All of this takes place on $\R^4$. 
 
 In our construction the Polyakov loop plays the role of an 
adjoint Higgs field. If the Polyakov loop is non-trivial, in the
sense that it has   non-degenerate eigenvalues in the weak coupling
domain, then it will lead to abelianization. This happens in 
QCD(adj) and in deformed Yang-Mills. In both cases $P=C$, where 
$C$ is the clock matrix. The difference as compared  to the adjoint
Higgs mechanism on $\R^4$ is that on $\R^{3} \times S^1$, the 
long distance theory is defined on $\R^{3}$. 

\begin{figure}[tbp] 
\begin{center}
\includegraphics[width=0.9\textwidth]{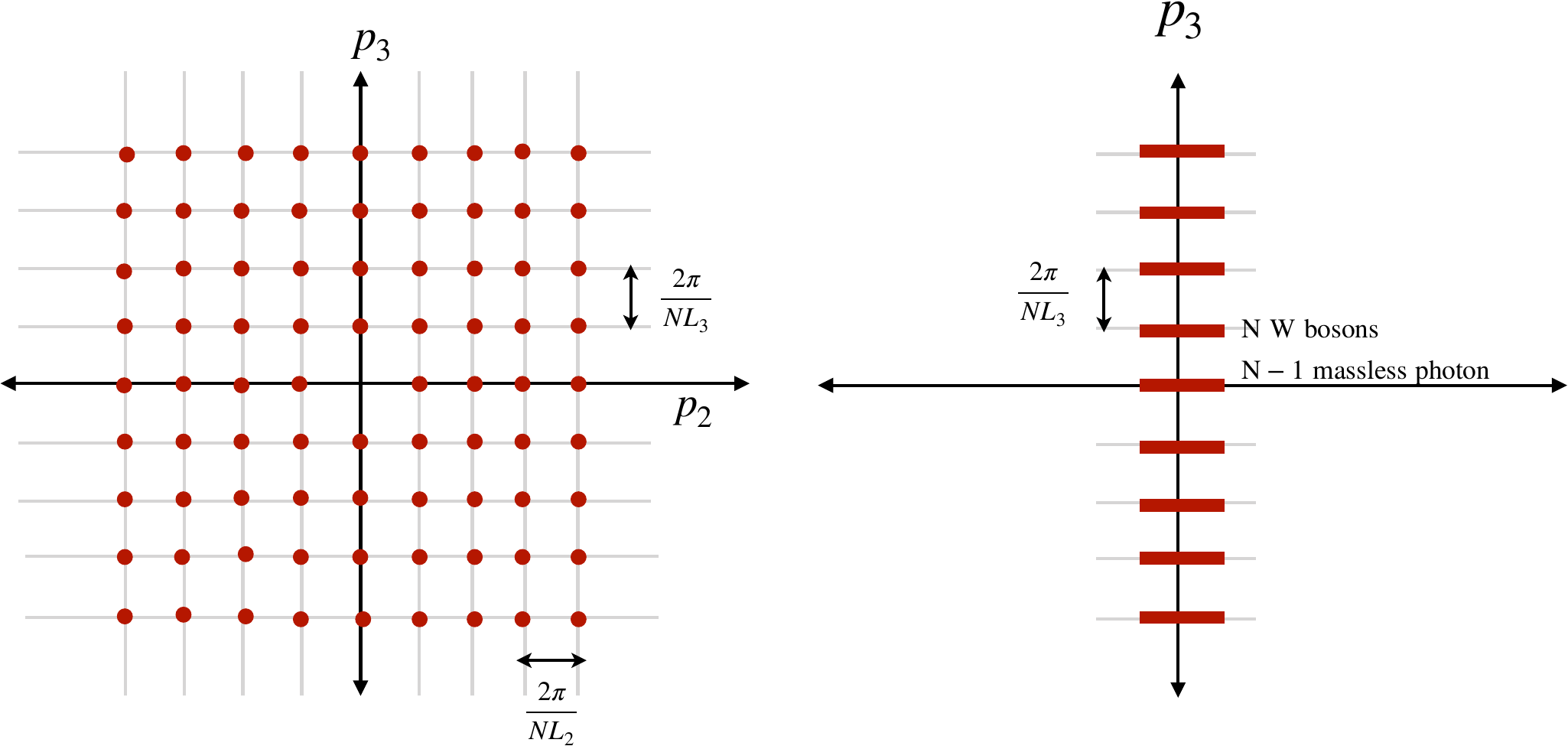} 
\end{center}
\caption{(Left) Spectrum of gluons in the presence of a two-stage adjoint 
Higgs mechanism based on non-commuting clock and shift matrices. No massless
modes survive. 
(Right) Spectrum of gluons for a one-stage adjoint Higgs mechanism with 
the clock matrix acting as a Higgs field. $N-1$ massless gluons remain. 
In the $L_2 \rightarrow \infty$ limit, the levels in the spectrum on the 
left become degenerate and we obtain the spectrum on the right. }
\label{fig:stage}
\end{figure}

  We now further compactify theory on $\R^{3} \times S^1$
to $\R^2\times (S^)_2 \times (S^1)_3$ with $L_2\gg L_3$. The
't Hooft flux $n_{23}=1$ gives rise to a two-stage adjoint Higgs 
mechanism with non-commuting Polyakov loops $P_3$ and  $P_2$ 
  \begin{align}
SU(N) \; \xrightarrow [ {\rm scale} \; L_3 ] 
  {P_3\, \text{adjoint Higgs}}  \;
    \underbrace{U(1)^{N-1}}_{\rm 3d \;  EFT}  \;
 \xrightarrow [ {\rm scale} \; L_2 ] 
  {P_2\,\text{adjoint Higgs}} \;
  \underbrace{\Z_N}_{\rm 2d \; EFT}, \qquad L_3 \ll L_2\, . 
\end{align} 
We first consider the effect of the Polyakov lines on the spectrum 
of the gluonic degrees of freedom \eqref{gluonspectrum} 
\begin{equation}
 M_{\bm{p},\bm{k}}^2
      =   \left({2\pi\over L_3}\right)^2 \left(k_3+ \frac{p_3}{N} \right)^2
        + \left({2\pi\over L_2}\right)^2 \left(k_2+ \frac{p_2}{N} \right)^2\, . 
\label{gluonspectrum2}
\end{equation}
We observe that $(N-1)$ gluons remain gapless in perturbation theory
as $L_2 \rightarrow \infty$, and that the $W$-bosons come in $N$-fold 
degenerate multiplets, as shown in Fig.~\ref{fig:stage}. At large but
finite $L_2$ the gapless gluons acquire a small mass $\frac{2 \pi}{NL_2}$ 
and the degeneracy among gluons is lifted.  

 This implies that the $N-1$ gapless photons on $\R^3$ acquire a 
classical mass (instead of the non-perturbative mass generated by 
monopoles)  
\begin{equation}
 M_{{p_2}, {k_2}}^2
      =   \left({2\pi\over L_2}\right)^2 \left(k_2+ \frac{p_2}{N} \right)^2\, . 
\label{photons2}
\end{equation}
Here, $k_2$ is Kaluza-Klein label, and $p_2$ is the (fractional) 
momentum label.  

 How does this translate to the dual photon description? First, recall 
that  the 1-form symmetry of the Yang-Mills theory on $\R^4$ decomposes 
on $\R^3 \times S^1$ as
\begin{align}
(\Z_N^{[1]})_{4d} \xrightarrow[]{ {\rm on}\; {\R^3 \times S^1}} 
  (\Z_N^{[1]})_{3d} \times  \Z_N^{[0]}\, , 
\end{align}
where $\Z_N^{[0]}$ is the zero form part of the center-symmetry. 
Ref.~\cite{Cherman:2016jtu} pointed out that the gauge invariant 
definition of the photon field is  $F_{\mu \nu, j} = \sum_{l=0}^{N-1} 
\omega^{j l} \tr (P_3^l F_{\mu \nu})$, and that the dual photon is 
$d \sigma_j \propto  * F_{\mu \nu, j}$. Since $\Z_N^{[0]}:P_3\rightarrow 
\omega P_3$, this symmetry cyclically rotates the dual photons as  
$\sigma_j \rightarrow \sigma_{j+1}$,  where  we denote the photon in 
an $N$-component notation $\bm \sigma = (\sigma_1, \ldots,  \sigma_N)$.  
The eigenstates of the $\Z_N^{[0]}$ symmetry are identical to the mass
eigenstates of the dual photon \cite{Unsal:2008ch}
\begin{align}
  \widetilde  \sigma_p =\frac{1}{\sqrt N} \sum_j \omega^{jp} \sigma_j\, . 
\end{align}
In this basis of $\Z_N^{[0]}$ eigenstates the 't Hooft twisted boundary
condition is
\begin{align}
 \widetilde \sigma_p (x_0, x_1, x_2+ L_2) =   
  e^{ \im \frac{2 \pi}{N} p}  \widetilde \sigma_p (x_0, x_1, x_2)  \; .
 \label{tbc}
\end{align}
Hence, the mode decomposition for the $\widetilde \sigma_p$ field can 
be written as (changing $p$ to $p_2$ for convenience)
\begin{align}
  \widetilde \sigma_{p_2} (x_0, x_1, x_2) = \sum_{k_2 \in \Z}    
  \widetilde \sigma_{p_2, k_2} (x_0, x_1) 
  e^{ \im \frac{2 \pi}{ L_2} ( k_2 + \frac{p_2}{N}) x_2} \, , 
\end{align}
producing the same spectrum as in \eqref{photons2} from a 2d 
QFT perspective. 

 We are now in a position to analyze the effective theories that
govern the two-stage Higgs mechanism. On $\R^3 \times S^1$ with 
a center-symmetric Polyakov loop $P_3$ the long-distance theory 
is described by a $(N-1)$ component magnetic Coulomb gas, where the
effective action is given in \eqref{master}. This theory has a 
non-perturbative mass gap $m_g=\ell_D^{-1}$ given in \eqref{Debye}.
Note that $m_g$ is non-perturbatively small, and $\ell_D$ is very 
large. As we compactify the theory further down to $\R^2 \times 
(S^1)_2 \times (S^1)_3$, there are two distinct regimes depending
on the relative magnitude of $L_2$ and  $\ell_D$. The two limits are
\begin{align} 
\label{3dreg} 
 L_2 \gg \ell_D  \gg L_{3}   \qquad {\rm 3d \; EFT} \, ,    \\
\label{2dreg} 
\ell_D  \gg  L_{2} \gg L_{3} \qquad {\rm 2d \; EFT} \, . 
\end{align}  
The 3d EFT describes confinement by semi-classical monopoles, and 
the 2d $\Z_N$ gauge theory confines because of 2d center-vortices.
For $L_s\sim \ell_D$ the two pictures are adiabatically connected.
The central question is how the 3d effective action \eqref{master} 
based on monopoles turns into to the deformed TQFT action in 2d. 

 
 Recall that in the regime $L_2 \gg \ell_D$ the mass gap is generated 
non-perturbatively by monopole-instantons. In the opposite regime, $L_{2}
\ll  \ell_D$, the mass gap is classical, and due to the presence of 
the 't Hooft flux. In the latter regime the twisted boundary
condition on the $\bm \sigma$ field lift the $N-1$ continuous moduli,
and $N$ discrete classical vacua  $\bm \sigma_{0, k} = \frac{2\pi}{N} 
{\bm \rho} k$ remain. These vacua are the critical points of the magnetic
Coulomb gas on $\R^3$, and the vacuum structure of the theory remains 
unchanged as the origin of the mass gap changes from non-perturbative 
to classical.


 In the regime $L_{2} \ll \ell_D$ the classical mass of the dual photons 
is a multiple of $\omega= \frac{2\pi}{NL_2}$. We add the corresponding
mass operator in the relevant vacuum. Dimensionally reducing the EFT 
on $\R^3 \times (S^1)_3$ down to  $\R^2 \times (S^1)_2 \times (S^1)_3$ 
we obtain 
\begin{align}
S_{\rm 2d}  =   
  L_{2} \int_{\R^2 } {g^2\over 8\pi^2 L_3}   \left( |\diff  \bm{\sigma} |^2  
 + \omega^2    \min_k  \left|\bm{\sigma} 
       - \frac{2 \pi}{N}  {\bm \rho} k  \right|^2 \right)
 - 2 \zeta_m \sum_{a=1}^{N} 
    \cos\left(\bm{\alpha}_a \cdot \bm{\sigma}+{\theta \over N}\right) \,  .
\label{QFT-master}
\end{align} 
This is a remarkable formula that incorporates many interesting insights. 
In particular, the theory based on abelianized $U(1)^{N-1}$ gauge theory 
is equivalent, in the long distance regime, to 2d deformed $\Z_N$
TQFT. To see this, note that at large $\omega^2$, the ground state is 
localizes on one of the $N$ gapped critical points $\bm{\sigma}_k^*=
{2\pi\over N}k \bm{\rho}, k=0, \ldots, N-1  $  along the Weyl vector
\eqref{branch}. Using the ansatz ${\bm\sigma} =  {\bm \rho}\varphi$, 
this leads to a 2d field theory
 \begin{align}
S_{\rm 2d}=   
 L_2 \int_{\R} {g^2 {\bm \rho}^2   \over 8\pi^2 L_3}   
   \left( |\diff \varphi  |^2  
   + \omega^2    \min_k  \left| \varphi - \frac{2 \pi}{N} k  \right|^2 \right)
   - 2 \zeta_m  N \cos\left(\varphi +{\theta \over N}\right) \,  . 
  \label{QFT-master2d}
 \end{align}
The first two terms in this EFT can be recast as a $\Z_N$ TQFT in the deep 
infrared, and the last term is just the center-vortex operator
\begin{align}
\label{TQFT1-2d}
S_{\rm 2d}  =  &  
 \frac{iN}{2 \pi} \int_M \varphi \, da  
 - 2\zeta_{\rm v} \int_M d^2 x \, \cos(\varphi + \theta/N) \, , 
\end{align} 
where we  identified $\zeta_{\rm m} L_2 N   \equiv \zeta_{\rm v}$ as 
the fugacity of the vortex gas. The monopole operator becomes 
\begin{align}
e^{-S_I/N} e^{\im \bm{\alpha}_a \cdot \bm{\sigma} 
    + \im \frac{\theta}{N} } 
 \rightarrow L_2 e^{-S_I/N} e^{\im \bm{\alpha}_a \cdot \bm{\sigma_k}^* 
   + \im \frac{\theta}{N} } = L_2 e^{-S_I/N} e^{\im  \frac{k}{N} 
   + \im \frac{\theta}{N} }\, ,
   \label{montovor}
\end{align}
which is the center-vortex operator in 2d. At low energy,  the partition function localizes to $N$ critical points of the dual photon field, and   
we obtain  
\begin{align} 
Z_{\rm 2d}  (\theta)&=
 \sum_{k}   e^{ V_{\R^2}
 2 \zeta_{\rm v}   \cos\left( \frac{2 \pi k+  \theta}{N}\right) }\, ,
\label{QFT2-master}
\end{align}
corresponding to the semi-classical center-vortex theory in 2d. 

 We can further compactify $\R^2 \times (S^1)_2 \times (S^1)_3$ to 
$\R\times (S^1)_1\times (S^1)_2 \times (S^1)_3$. This reduces 
\eqref{QFT-master} to quantum mechanics. Retaining only the zero 
mode of the $\bm{\sigma}$ field we obtain the 1d effective action
\begin{align}
S_{\rm 1d}=   L_{1} L_2  \int_{\R} {g^2\over 8\pi^2 L_3}   
  \left( |\diff_\tau \bm{\sigma} |^2  
   + \omega^2 \min_k  
   \left|\bm{\sigma} - \frac{2 \pi}{N}  {\bm \rho} k  \right|^2 
    \right)
 - 2 \zeta_m \sum_{a=1}^{N} \cos\left(\bm{\alpha}_a \cdot \bm{\sigma}
    +{\theta \over N}\right)\,. 
\label{QM-master}
\end{align}
It is worthwhile noting that this action, obtained for $n_{23}=1$, 
looks rather different from the one obtained by quantum mechanical 
reduction of the theory with $n_{12}=1$, \eqref{master-new}. Yet, 
both flow to the same deformation of the $\Z_N$ TQFT.

\subsection{Monopole interactions and flux fractionalization}
\label{sec:fluxfrac}

 How did the magnetic Coulomb gas in 3d turn into a dilute vortex
gas in 2d? To answer this question, we have to understand how
monopoles can generate center-vortices.
 
 Using \eqref{QFT-master} we can calculate the interaction between 
monopoles in this regime. In the context of naive dimensional
reduction from a 3d Coulomb gas to 2d the $\pm 1/r$ interactions
between magnetic objects turns into $\mp \log (r)$ in 2d, see, 
for example \cite{Anber:2011gn,Anber:2013doa}. However, this is 
not what happens in the compactification studied in this work,
which can be viewed as a \emph{twisted dimensional reduction}. In 
particular, we know that there are center vortices in 2d, and that 
the interaction between them must be short ranged, since all the 
degrees of freedom are gapped. This implies that center vortices
are analogous to gauged vortices in the abelian Higgs model (see 
\cite{Manton:2004tk}), but different from global vortices, such as 
the ones that arise in a neutral superfluid.

  The compactification we are considering involves a 't Hooft flux, 
which renders dual photons massive classically. We can inspect the
correlation functions of monopoles in the perturbative vacuum 
on $\R^2 \times (S^1)_2 \times (S^1)_3$ in the regime \eqref{2dreg},
in which the long-distance theory becomes 2d. We find   
\begin{align} 
\langle {\cal M}_i({ { x}}) {\cal M}_j({ 0}) \rangle_{\rm free, \; 2d} 
 & = \langle \rme^{\im\bm{\alpha}_i \cdot \bm{\sigma}({ x})} 
             \rme^{\im\bm{\alpha}_j \cdot \bm{\sigma}({ 0})} \rangle \, . 
\end{align}
Assuming that $|{ x}| > \frac{NL_2}{2 \pi}$ where ${ x} $ 
is the separation on $\R^2$, the interactions are given by
\begin{align} 
 V({ x}) &=   \frac{ A_{ij}}{L_2}   
    \int \frac{d^2p}{(2 \pi)^2} 
       \frac{e^{\im p.{ x}}}{p^2 + \omega^2}  
   =    \frac{ A_{ij} }{L_2}     
        \frac{1}{2\pi} K_0 ( \omega |{ x} |)   \cr
  & \approx    \frac{ A_{ij}}{L_2}   
    \frac{1}{4\pi} \left( \frac{NL_2}{|{ x}|} \right)^{\frac{1}{2}} 
     e^{ - \frac{2 \pi}{NL_2} |{ x}|} \quad\quad  
     \left(\omega |{ x} | \rightarrow \infty\right)\,  ,   
  \label{int-exp} 
\end{align}
where we have used the asymptotic behavior of the modified Bessel 
function in the last step.\footnote{
We can work without the assumption $|{x}| > (NL_2)/(2 \pi)$. In that 
case, we have to work with the full 3d Green function on $\R^2 \times 
S^1$ with twisted boundary condition on $S^1$. This gives
\begin{align}
\int \frac{d^2p}{(2 \pi)^2} \frac{e^{\im p.x}}
 {p^2 + (\frac{2 \pi}{L_2N})^2 }  \rightarrow  \sum_{p_3=1}^N 
 \sum_{k_3 \in \Z} \int \frac{d^2p}{(2 \pi)^2} 
 \frac{e^{\im p.x}}{p^2 + 
  (\frac{2 \pi}{L_2} (k_3+ \frac{p_3}{N}))^2 }  = 
 {\sum_{q_3 \in \Z}}^{'} \int \frac{d^2p}{(2 \pi)^2} 
 \frac{e^{\im p.x}}{p^2 + (\frac{2 \pi q_3}{NL_2} )^2 }\, . 
\end{align} 
In the last step the prime indicates the absence of zero mode at 
finite $NL_2$.  In the limit  $NL_2\rightarrow \infty$ or $|x| 
\ll NL_2$ we obtain 3d Coulomb interaction $1/(4\pi |x|)$, while 
at large distance we get an exponential fall-off $\exp[-(2\pi)|x|
/(NL_2)]$.}
The important point is that the interaction becomes short-range,
and that the magnetic flux does not spread on $\R^2$. It is localized 
in a tube of size $\omega^{-1} = \frac{NL_2}{2 \pi}$. Note that the 
monopole size is $(m_W)^{-1} = \frac{NL_3}{2 \pi}$. To summarize, 
\begin{align} 
{\rm monopole \; core \; size\;  on \; } \R^2 \times (S^1)_2:
    & \qquad   r_{\rm m} = (m_W)^{-1} =   \frac{NL_3}{2 \pi} \, , \cr 
{\rm center-vortex \; size \;  on\; } \R^2: \quad\;
    &\qquad   r_{\rm v}  =   \frac{NL_2}{2 \pi}    \, . 
\label{sizes}
\end{align}

\begin{figure}[tbp] %
\begin{center}
\includegraphics[width=0.9\textwidth]{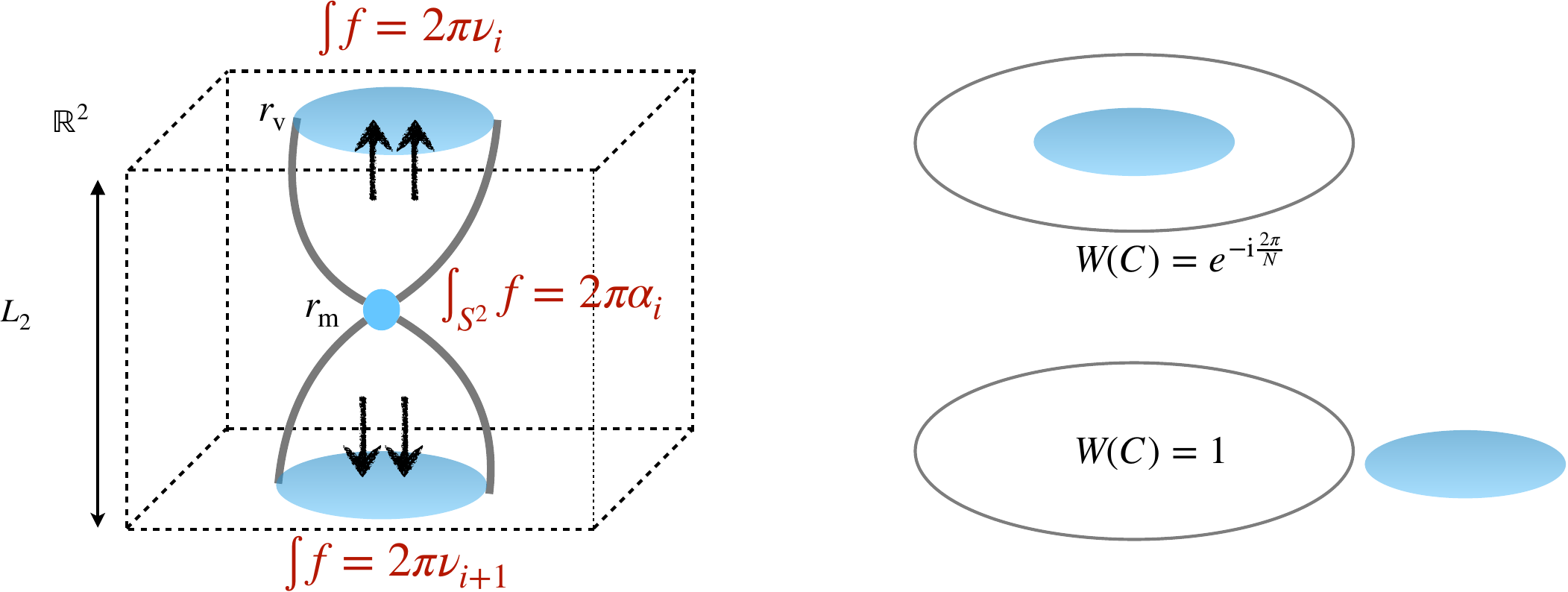} 
\end{center}   
\caption{(Left)  A monopole centered at $x_2^*$ in $\R^2 \times 
(S^1)_2$ ($(S^1)_3$ not shown). Due to the 't Hooft flux $n_{23}=1$ 
the monopole flux (which is spherically symmetric at short distances)
collimates into a tube of characteristic size $r_{\rm v}$.      
(Right) A Euclidean 2d observable on $\R^2$ seed this flux as 
a center-vortex due to its statistics with Wilson loop.}
\label{fig:mon-vor}
\end{figure}

 We can inspect the flux emanating from a monopole centered at 
$x_2^*$ in $\R^2 \times (S^1)_2$. We will assume that $N$ is of order 
one.\footnote{
The reason we do so here is related to subtleties arising in the large
$N$ limit. Notice that both the microscopic size of a monopole as well 
the center-vortex size depends on $N$. This is not an accident. Even 
at small $L_3, L_2$, if $N$ is large enough, one can no longer use 
weak-coupling semi-classical methods. This is related to large-$N$ 
volume independence or (twisted) Eguchi-Kawai reduction 
\cite{Eguchi:1982nm,GonzalezArroyo:1982hz,Kovtun:2007py}.}
The flux is  spherically symmetric at distances much smaller than 
$L_2$.  If we consider a sufficiently large $S^2$ surrounding the 
monopole we find that the flux fractionalizes into two tubes.  
${\bm \nu}_i$ units of magnetic flux pass through the vicinity 
of the North Pole, and $-{\bm \nu}_{i+1}$ units of magnetic flux 
thread the vicinity of the South pole, see Fig.~\ref{fig:mon-vor}. 
The total flux is $ {\bm \alpha}_{i} = {\bm \nu}_i - {\bm \nu}_{i+1}$, 
equal to the monopole charge. 

  Remarkably, a Euclidean  observable on 2d slice $\R^2$, will 
see either ${\bm \nu}_{i+1}$ or ${\bm \nu}_i$ units of flux. The 
flux can be detected using a large Wilson loop in the fundamental
representation (larger than $NL_2/2\pi$). Consider a Wilson loop 
on $\R^2$, $W(C)= \frac{1}{N}\,\tr\, e^{ \im \oint  a}$. In the
abelianized theory, only photons in the Cartan subalgebra survive 
and we can write  
\begin{align}
W(C) =  \frac{1}{N}\, \tr \, e^{ \im \oint  a} 
 \xrightarrow{SU(N)\to U(1)^{N-1}} 
 \frac{1}{N} \sum_{j=1}^{N} e^{ \im \oint {\bm \nu}_j\cdot a}  = 
 \frac{1}{N} \sum_{j=1}^{N} e^{ \im \int_D {\bm \nu}_j\cdot f}   \; ,
\end{align}
where $f= \diff a$ and $C= \partial D$. Since  ${\bm \nu}_i.
{\bm \nu}_j= \delta_{ij}-\frac{1}{N}$, the flux $f= 2\pi
{\bm \nu}_i$ (for $x_2>x_2^{*}$) or $f= 2\pi{\bm \nu}_{i+1}$ 
(for $x_2 < x_2^{*}$) leads to 
\begin{align} 
W(C) =   e^{ - \im \frac{2 \pi}{N} \Theta_D(x)}   =   
 \left\{  \begin{array}{cl}   
   e^{ - \im \frac{2 \pi}{N}}  & \qquad {\rm vortex} \in D, \cr 
                 1             & \qquad {\rm vortex} \notin D, 
\end{array}   \right.
\label{Vortex-def}
\end{align} 
where $\Theta_D(x)$ is the support function, which is equal 
to $1$ for $x\in D$ and $0$ otherwise. Euclidean 2d operators 
see the flux emanating from the monopole as a center-vortex. We 
refer to this phenomenon as $\emph{flux fractionalization}$. 
The position of the monopole in the compactified direction 
$x_2^*$ can be interpreted as an internal moduli parameter 
of the vortex. 

 When $L_2$ is asymptotically large, then a monopole at position 
$x \in \R^3$ with flux (charge) $2 \pi \alpha_j$ is detected by 
abelian Wilson loop with charge ${\bm \nu}_i$ as 
\begin{align}
W_i(C) = e^{\im {\bm \nu}_i  \oint a} 
   =    e^{\im {\bm \nu}_i \int_D f } 
   =  e^{\im  \half {\bm \nu}_i \alpha_j \Omega_D(x) } 
   =  e^{\im  \half (\delta_{ij} - \delta_{i, j+1}) \Omega_D(x) } ,
  \label{mon-phase0}
\end{align}
where $\Omega_D( x)$ is the solid angle for the oriented surface 
$D (\partial D = C)$ subtended at a point $x \in \R^3$. For an 
asymptotically large area $A$, filling a plane $\R^2$, the solid 
angle is equal to half the solid angle of a sphere. If $x$ is below 
$C$ the solid angle is $2 \pi$, hence 
\begin{align}
\lim_{D \rightarrow \R^2}  W_i(C)  = 
  e^{\im  \pi  (\delta_{ij} - \delta_{i, j+1}) } 
  \label{mon-phase}
\end{align}
In a grand-canonical ensemble of monopoles it is the sum over 
these phases that generates the area law of confinement on $\R^3$. 
The main point is that in both the monopole and center-vortex 
regimes these phases are present, both are generated by monopoles, 
and both lead to the area law of confinement.

 \section{Outlook: Metamorphosis} 

 In this work we studied the connection between the two main 
semi-classical mechanism for confinement that have been discussed
in the literature, monopoles and center-vortices. For this purpose
we consider weak coupling calculations on $\R\times T^2\times S^1$.
In \S.\ref{sec:sym} and \ref{sec:T2small} we show how to connect the 
monopole EFT on $\R^3\times S^1$ with the vortex EFT on $\R^2\times
T^2$ via a path through a quantum mechanical regime.

 On $\R^3\times (S^1)_3$ with a center-symmetric holonomy the 
EFT contains dual photons and monopole operators. The monopole
fugacity is controlled by the action $\frac{S_I}{N}$, and there 
are long-range interactions between them. The proliferation of 
monopole-instantons generates a mass gap (magnetic Debye length 
$\ell_D$) as well as confinement and fractional theta dependence. 
These phenomena take place in a semi-classically calculable regime 
for small $S^1$, which is adiabatically connected to $\R^4$.  

In \S.\ref{sec:asym}, we study the theory on 
  $\R^2 \times (S^1)_2 \times (S^1)_3 $ with a 't Hooft flux 
$n_{23} =1$, at fixed small $L_3$ but with a varying $L_2$. 
 For 
$ L_2 \gg \ell_D$, the local dynamics is same as the theory on  $\R^3\times (S^1)_3 $.
However, for  $ L_2 \ll \ell_D$, we find that at 
large distances on $\R^2$ monopoles transmute to 2d center-vortices. We derived a long-distance EFT for the vortices starting with the EFT of monopoles. 
These vortices have short-range interaction, and non-trivial mutual
statistics with the Wilson loop due to flux fractionalization. Hence,
vortices lead to  confinement and fractional theta dependence  
in a semi-classically calculable regime on $T^2 \times \R^2$. 
Upon further compactification of the center-vortex theory to quantum
mechanics on small $\R\times T^3$, center-vortices (monopoles) become 
1d instantons  with short (long) range interaction interpolating 
between the $N$-perturbative vacua. This leads to confinement and 
fractional theta dependence.  

 In \S.\ref{sec:asym} and \S.\ref{sec:analogy},  
 we  also take a more microscopic view, and ask
how the semi-classical vortex configuration can arise from a 
monopole-instanton. Kraan and van Baal \cite{Kraan:1998pm,
Kraan:1998sn} showed how four-dimensional instantons in the presence 
of a Polyakov line along a compact direction fractionalize, and give 
rise to monople-instantons. Here we show how monopole-instantons in 
the presence of a 't Hooft flux give rise to center-vortices.  The problem for $SU(2)$ reduces to finding electric potential of a point charge  between conducting parallel plates,  which also exhibits a transmutation from  $1/r$ algebraic decay  between short and  intermediate distances  to exponential decay 
$e^{ -\frac{2 \pi}{2L_2}  \rho} $ at  distances larger than interplate 
separation $L_2$.\footnote{Chapter 17 of Schwinger's Classical Electrodynamics \cite{Schwinger} is devoted to this electrostatic problem.}

It is remarkable to notice that  all these objects are ultimately descendants  of  a 
four-dimensional
instanton probed by semi-classical fluxes or Polyakov lines. It 
is natural to ask whether one can extend these ideas directly 
to the four-dimensional theory, for example by finding suitable 
parametrizations of the instanton moduli space, and obtain a deeper understanding of the inner structure of an instanton.  

\noindent
{\bf Note added:} During the completion of this work, we learned 
about Ref.\cite{Hayashi:2024yjc}, which uses $\mathbb{R}^2 \times 
T^2$ with $n_{23}=1$. Our construction in \S.\ref{sec:sym} and  
\S.\ref{sec:T2small} uses $n_{12}=1$, and gives a complementary 
perspective on monopole-vortex transmutation. Our \S.~\ref{sec:asym} 
and Appendix \ref{sec:analogy} are motivated by this work. In contrast
with Ref.~\cite{Hayashi:2024yjc} we derive the 2d effective theory in 
terms of dual photon variables of the 3d monopole theory. After
the present work was completed, Wandler published a numerical 
study \cite{Wandler:2024hsq} which addresses some of the issues 
studied in our work.

\noindent
{\bf Acknowledgements:}
We thank  Antonio Gonzalez-Arroyo, Margarita Garcia Perez,  
Yuya Tanizaki, Yui Hayashi, David Wandler, Erich Poppitz,  
Aleksey Cherman, and Mendel Nguyen for useful discussions. 
The work is supported by U.S. Department of Energy, Office 
of Science, Office of Nuclear Physics under Award Number 
DE-FG02-03ER41260.

\appendix
\section{From charge-$N$ 2d Abelian Higgs model to TQFT} 

 Consider the charge-$N$ Abelian Higgs model with Lagrangian 
\begin{equation}
\mathcal{L} = {1\over 2e^2}|f_{12}|^2
  + |(\partial_\mu +{\rm i} N a_\mu)\Phi|^2
  + \lambda(|\Phi|^2-v^2)^2-{{\rm i} \theta\over 2\pi}f_{12}, 
\label{eq:Lagrangian_AbelianHiggs}
\end{equation}
In the classical vacuum, $\Phi = v e^{{\rm i} \xi} $. In the following
we set $v=1$, and ignore the fluctuations of the modulus. The kinetic 
term for the scalar field can be expressed as
\begin{align} 
|\partial_{\mu} \xi +N a_{\mu}|^2 \, . 
\label{original}
\end{align} 
We can dualize this Lagrangian by introducing a vector field $B_{\mu}$ 
and writing an auxiliary Lagrangian
\begin{align} 
 \mathcal{L}_{\rm aux} =  
 \frac{1}{4} B_{\mu}^2  
  + {\rm i}  \epsilon_{ \mu \nu} B_{\mu} (\partial_{\nu} \xi 
   +N a_{\nu})\, . 
\label{master2}
\end{align}
Integrating out $B$ first amounts to setting $B_{\mu} = 2 {\rm i}
\epsilon_{\mu \nu} ( \partial_{\nu} \xi +N a_{\nu})$ and we recover 
the original Lagrangian, \eqref{original}. Instead, if we integrate
out $\xi$ first, we obtain $ \epsilon_{\mu\nu} \partial_{\nu} 
B_{\mu} =0$. Hence, we set $B_{\mu}= \frac{1}{2 \pi} \partial_{\mu} 
\varphi$ where $\varphi$ is a $2 \pi$ periodic field. In terms of 
the dual variable $\varphi$ the Lagrangian can be rewritten as
\begin{equation}
\mathcal{L} = {1\over 2e^2}|d a |^2
   +  \frac{1}{ (4 \pi)^2} |  d \varphi|^2 
   +   \frac{iN}{2 \pi}  \varphi d a  
   -   {{\rm i} \theta\over 2\pi}f_{12}\, .  
\label{eq:dualLag}
\end{equation}
The dual theory has a $U(1)$ gauge field, but there is no field which is 
charged under this $U(1)$. Therefore, the gauge field only enters via its 
field strength. It is known in the context of the standard abelian Higgs 
model that the  effect of the gauge field is to generate a mass term for  
$\varphi$ \cite{Coleman:1976uz}, and that the IR theory is gapped. 
However, in the charge-$N \geq 2 $ Abelian Higgs model the IR theory 
is not completely trivial. At this stage, we can proceed in two different
ways, leading to overlapping and complementary results. 
 
In the low energy limit $ \frac{E}{e} \ll 1$ or equivalently, 
$e \rightarrow \infty$, we can drop the gauge kinetic term. The 
classical equation of motion for the gauge field $a$ leads to the 
constraint $d \varphi=0$, i.e. $\varphi$ becomes non-dynamical. To 
all orders in perturbation theory, the  effective 2d theory at large 
distances is the topological $\Z_N$ gauge theory (TQFT). The action of   
the $\Z_N$ TQFT can be expressed as \cite{Banks:2010zn,Kapustin:2014gua}
\begin{equation}
\label{TQFT-general}
    S_* = \frac{iN}{2 \pi} \int_M \varphi \, da,
\end{equation}
The fundamental question at this stage, then, is whether this infrared 
limit is stable or unstable.  We will see that with the inclusion of 
non-perturbative effects (instantons, which are vortices in this set-up)  
is crucial in this context.

Before answering this question, we note that in the TQFT limit the theory
no longer contains the  mass of the $\varphi$ fluctuations. In order to
establish the relation with  the statistical field theory that describes 
the vacuum of the model, it is useful to keep the fluctuating massive 
scalars in the IR description. 

  We can keep the mass of the scalar $\varphi$ using the following 
procedure. Since there is no charged matter field in the dual 
description, the appearance of the gauge field is through the field
strength $f= d a$. If we view the theory on ${\mathbb R}^2$ as the
decompactification limit of the theory on some $M_2$, we can replace 
the integral over $a$ with an integral over $f$ with the inclusion 
of a sum over all flux sectors, i.e, $\int Da \rightarrow
\int Df \sum_{\nu \in \Z} \delta \left( \nu - \frac{1}{2\pi} 
\int f \right)$. The latter integral can further be replaced with  
$\int Df  \sum_{n  \in \Z}e^{ {\rm i} n \int f }$ by using the Poisson
resummation formula. As a result, we can integrate out $f$ at a finite 
value of $e^2$ to obtain
\begin{align}
&\int Da \;   {\rm exp} \left[
  -{1\over 2e^2}\int |f|^2 
  + \frac{{\rm i}}{2\pi}   N \int \varphi f  
  +  \frac{{\rm i}}{2\pi}  \theta \int  f    \right]   \cr
&=     \int Df  \;   \sum_{n  \in \Z} \;  { \rm exp} \left[ 
   - {1\over 2e^2}\int |f|^2 
   + \frac{{\rm i} N }{2 \pi}    
     \int \left( \varphi + \frac{ \theta+ 2 \pi n }{N} \right) f  
     \right] \cr
&= C \;   {\rm exp}\left[ 
   - \frac{1}{2} \frac{e^2N^2 }{4 \pi^2}  
        \int \min_{n \in \Z}  
    \left ( \varphi + \frac{ \theta + 2 \pi n}{N} \right)^2 \right]\, ,
\end{align}  
where $C$ is a constant. Therefore, the dual Lagrangian to all orders in
perturbation theory takes the form
\begin{align}
   \mathcal{L}= 
     \frac{1}{(4 \pi)^2}  |  d \varphi|^2 
 +   \frac{1}{2} \frac{e^2N^2 }{4 \pi^2} \min_{n \in \Z}  
        \left ( \varphi + \frac{ \theta + 2 \pi n}{N} \right)^2\, , 
\label{eq:dualLag2}
\end{align} 
i.e. $\varphi$ is free massive scalar.  Within the fundamental domain of 
$\varphi$, the theory has $N$  minima. Setting $\theta=0 $ these minima
are located at  
\begin{align}
     \varphi_*=   \frac{ 2 \pi n}{N}, \;\; n=0, 1, \ldots, N-1\, .
\label{min}
\end{align}
The massive theory, \eqref{eq:dualLag2}, in the deep infrared, reduces 
to TQFT. Indeed, the minima \eqref{min} are consistent with the TQFT
action, for which the equation of motion of $a$ yields $d \varphi=0$, 
and the possible values of $\varphi$ are given by the same set of 
discrete points. 

The inclusion of vortices amounts to deforming  \eqref{TQFT-general} or
\eqref{eq:dualLag2} as
\begin{align}
\label{def-QFT-AHM1}
  S  &= \frac{iN}{2 \pi} \int_M \varphi \, da 
    -   2 \zeta_{\rm v} \int_M  \cos{ (\varphi + {\theta}/{N}}) \\
  S  &= \int_M  \frac{1}{(4 \pi)^2}  |  d \varphi|^2 
    +   \frac{1}{2}  m_\varphi^2 ``\varphi^2"  
    -   2 \zeta_{\rm v} \int_M  \cos ( {\varphi + {\theta}/{N}) }\, , 
\label{def-QFT-AHM2}
\end{align}
where the mass term $m_\varphi^2 ``\varphi^2"$ has to be understood as 
in \eqref{eq:dualLag2} and we shifted $\varphi$ to have the $\theta$ 
angle in the vortex term.  

There is however, a disadvantage to working with the deformed TQFT. 
This EFT does not account for the interaction between vortices, and 
treats them as non-interacting. This is justified because the interaction
deduced from \eqref{def-QFT-AHM2} by inspecting the vortex-vortex
correlators is
\begin{align}
    \langle e^{\im \varphi(x) } e^{ \pm \im \varphi(0) } 
      \rangle_{\rm free} 
    = e^{-V(x)} \quad \Longrightarrow \quad  V(x) 
    \sim \pm K_0(m_\varphi |x|) \sim \pm e^{-m_\varphi |x|}\, . 
\end{align}
The TQFT limit corresponds to $m_\varphi \rightarrow \infty$, and 
the interactions does indeed disappear. However, in our story, in 
order to explain how the long-range interaction between monopoles 
transmutes to the short-range interaction among vortices, it is very 
important to keep track of the interactions. More generally,  it is 
useful to keep track of the interactions between topological 
configurations when we map the Euclidean vacuum structure of a 
theory to statistical field theory. 

\section{An electrostatic analogy: Flux fractionalization}
\label{sec:analogy}

 In the case of $SU(N)$ gauge theory Hayashi and Tanizaki pointed 
out that the relation between monopoles in a 't Hooft flux background 
and center-vortices can be mapped to an electro-static problem \cite{Hayashi:2024yjc}. Here, we detail the simplest realization of this to $SU(2)$ gauge theory in detail.  The goal is to point out that the potential becomes short range, exponentially decaying away from the chain. This is in agreement with our findings \eqref{int-exp} obtained by evaluating monopole-correlators in $n_{12}=1$ background.

Consider an infinite line array of alternating charges,  $\pm$, separated 
from each other by the spacing $L_2$. We can compute the electrostatic 
potential (in cylindrical coordinates) at position $ \vec \rho \equiv 
(x_0, x_1)$ and $x_2$. Let us position  $\pm$ charges at $(x_2^{*}+ 2L_2 k$, 
$(x_2^{*} + 2L_2 (k+ \half))$, respectively, and shift the coordinates 
such that $x_2^{*}=0$  for convenience.\footnote{In the microscopic 
theory, the $+$ charge comes from the BPS monopole and $-$ charge arises 
from KK monopoles, both with $Q_T=\half$ topological charge, and 
$e^{\im \frac{\theta}{2}}$ theta angle dependence. It is important to 
note that  $-$ charge is {\it not} an anti-monopole. In the case of a 
general $SU(N)$ group twisted boundary condition imply the presence of 
a chain of monopoles $\ldots \alpha_1 \alpha_2 \ldots \alpha_N \alpha_1 
\alpha_2 \ldots \alpha_N \ldots$ repeating indefinitely, each monopole 
with $e^{\im \frac{\theta}{N}}$ theta dependence.} 

\begin{figure}[tbp] 
\begin{center}
\includegraphics[width=0.7\textwidth]{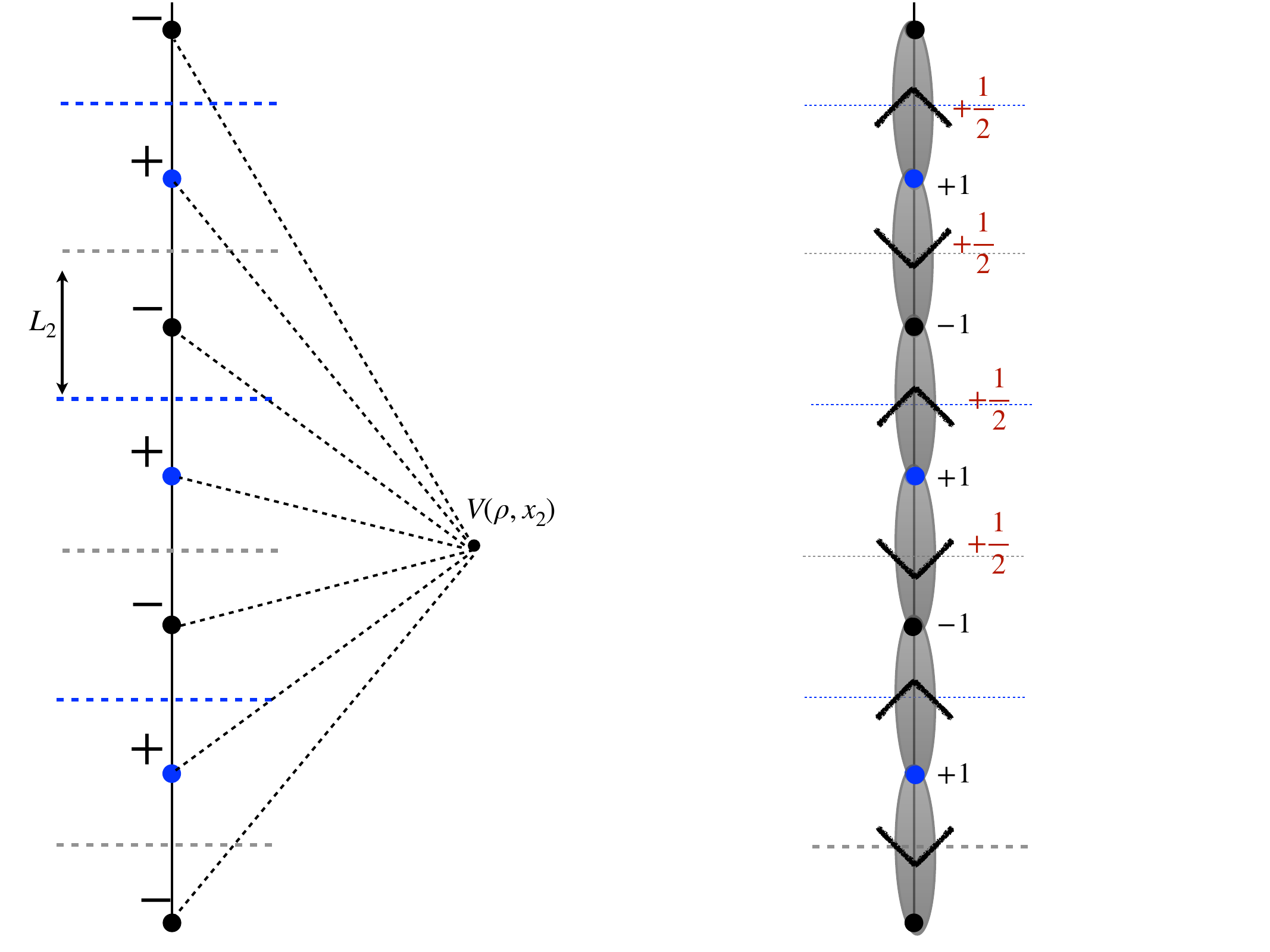} 
\end{center}
\caption{(Left) An alternating array of $\pm$ charges. The potential 
evaluated at $\rho, x_2$ decays exponentially for large $\rho$ as 
$e^{-\frac{2 \pi}{2L_2} \rho}$.  (Right) The flux of a charged particle 
collimates into a tube of thickness  $2 L_2/2\pi$.  The fractionalized 
flux  gives a contribution $W(C) = e^{\im \pi}$ to a Wilson loop on 
$\R^2$, leading to area law of confinement on $\R^2 \times T^2$. }
\label{fig:frac}
\end{figure}
 
This is an interesting problem, because each type of charge array $(\pm)$
individualy produces a $\mp \log \rho$ term at large distances.  But of 
course, due to the alternating charges, we expect $V(\rho, x_2) \rightarrow 
0$ as $\rho \rightarrow \infty$. But the question is how does the potential 
decay exactly, algebraically or exponentially?  In the analysis in the 
bulk of the paper, we reduced  the 3d Coulomb gas EFT down to a deformed 
TQFT in 2d. This implies  that the fall-off  must be exponential. We would 
like to see that the exponential decay of the potential emerges from a 
sum of the type $\sum_{k \in \Z} \frac{Q_k}{|r-r_k|}$.
 
The potential at some $(\rho, x_2)$ can be written as  
\begin{align}
V(\rho, x_2) =  \frac{1}{4\pi} \sum_{k \in \Z}  
       \frac{1}{\sqrt {(x_2 - 2L_2 k)^2 + \rho^2}} 
            -  \frac{1}{4\pi} \sum_{k \in Z} 
       \frac{1}{\sqrt {(x_2 - 2L_2 (k + \half) )^2 + \rho^2}}\, . 
\end{align} 
Clearly, $ V(\rho, x_2)$ is periodic with period $2 L_2$ in the $x_2$
direction, and $ V(\rho, x_2) = V(\rho, -x_2)$. Therefore, we can 
Fourier expand $ V(\rho, x_2)$  in the  orthogonal eigenbasis 
$\{ \cos( \frac{2\pi}{2L_2} m x_2),  \; m=[0, \infty) \}$ as
\begin{align}
 V(\rho, x_2)  = \sum_{m=0}^{\infty} V_m(\rho) 
    \cos\left(\frac{2 \pi}{2L_2} m x_2 \right) \, . 
\end{align} 
The Fourier coefficients $V_m(\rho)$ tells us how each mode behaves 
as a function of $\rho$, and can be found via inverse transform
\begin{align}
 V_m(\rho)  &=  \frac{1}{2 L_2} \int_{-L_2}^{+L_2}dx_2\,   
    V(\rho, x_2)   \cos \left(\frac{2 \pi}{2L_2} m x_2 \right),  \cr 
  &=  \frac{1}{2 L_2}    \left(1 - (-1)^m\right) 
    \int_{-\infty}^{+\infty} dx_2\,    
       \frac{1}{ 4 \pi \sqrt{  {(x_2)^2 + \rho^2}} } 
       \cos\left(\frac{2 \pi}{2L_2} m x_2 \right)\, .  
\end{align}
Remarkably, the Fourier coefficient of the zero mode $m=0$ vanishes, 
as well as all even modes, $m \in 2\Z^{+}$.  We could have anticipated 
this easily because the potential at $x_2 = L_3/2 + L_3\Z$ must vanish 
at any $\rho$ by symmetry. For $m \in 2\Z^{+}+1$, we obtain the Fourier 
coefficients
\begin{align}
 V_m(\rho)   
  &=  \frac{1}{2 \pi  L_2}    
     K_0 \left(\frac{2 \pi}{2L_2} m \rho  \right)\, . 
\end{align}
This implies that the potential has the form 
\begin{align}
 V(\rho, x_2)  =   \frac{1}{2 \pi  L_2}     
   \sum_{m=1,3, \ldots}     
       K_0 \left(\frac{2 \pi}{2L_2} m \rho  \right)
      \cos \left(\frac{2 \pi}{2L_2} m x_2 \right), 
\end{align} 
At short distances, close to the cores of monopoles, this potential 
behaves as $\pm 1/r$ as it should. At long distances, all modes 
decay exponentially fast for $\rho > \frac{2L_2}{2 \pi}$. Even the 
slowest decaying mode falls off as $e^{ -\frac{2 \pi}{2L_2}  \rho} $.
This is the same result as we obtained in \eqref{int-exp} from 
field theory. 

\bibliographystyle{utphys}
\bibliography{./QFT-Mithat.bib,refs}
\end{document}